# General Properties of Solutions to Inhomogeneous Black-Scholes Equations with Discontinuous Maturity Payoffs and Application

**Hyong-Chol O** [1] and **Ji-Sok Kim** [2]

[1, 2]*Faculty of Mathematics,* ***Kim Il Sung*** *University, Pyongyang, D. P. R. Korea*
[2] *Corresponding Author E-mail address:* kjisok@yahoo.com

*Abstract*: We provide representations of solutions to terminal value problems of inhomogeneous Black-Scholes equations and studied such general properties as min-max estimates, gradient estimates, monotonicity and convexity of the solutions with respect to the stock price variable, which are important for financial security pricing. In particular, we focus on finding representation of the gradient (with respect to the stock price variable) of solutions to the terminal value problems with discontinuous terminal payoffs or inhomogeneous terms. Such terminal value problems are often encountered in pricing problems of compound-like options such as Bermudan options or defaultable bonds with discrete default barrier, default intensity and endogenous default recovery. Our results are applied in pricing defaultable discrete coupon bonds.

## 1.  Introduction

Since the development of the famous Black-Scholes formula for option pricing in [3], the Black-Scholes equations have been widely studied and became one of the most important tools in financial mathematics, in particular, in pricing options and many other financial securities.

In this paper, we study general properties of the solutions to terminal value problems of *inhomogeneous Black-Scholes partial differential equations*. We provide the *solution - representations*, *min-max estimates* of the solutions, *gradient estimates* with respect the stock price variable, *monotonicity* and *convexity* of the solutions.

It is well known that standard European call and put options have pricing formulae (solution representation of a special terminal value problems of Black-Scholes equations) and their price functions have various monotonicity (with respect to stock price, exercise price, short rate, dividend rate, and volatility) and convexity (with respect to stock price and exercise price) [10, 13].

In 1973, Fischer Black and Myron Scholes [3] gave the call option pricing formula, so called Black-Scholes formula. More general formulae of the solutions to Black-Scholes equations with any terminal payoffs are described in [13, 17] and etc.

Merton [14] and Jagannathan [9] show that a call option's price is an increasing and convex function of the stock price under the assumption that the risk-neutralized stock price follows a proportional stochastic process (under which the pricing model becomes a simple Black - Scholes equation with zero short rate and dividend rate). Cox et al. [5] generalize this result and show that, under the same proportionality assumption of the stock price process, the price function of any European contingent claim, not just a call option, inherits qualitative properties of the claim's contractual payoff function [2]. (This means that the solution to Black-Scholes equation inherits such qualitative properties as monotonicity or convexity of the maturity payoff function.)

Bergman et al. [2] generalized the results of [5, 9, 14] into the case when the stock price follows a diffusion whose volatility not only depends on time but also is a differentiable function of and current stock price. They used PDE method to study gradient estimate and convexity preservation. In particular, the property of convexity preservation is related to the monotonicity of option price with respect to the volatility and widely studied. El Karoui et al. [6] obtained the same results as [2] with stochastic approach. Hobson [8] simplified the result of [6] using stochastic coupling. Janson et al. [11] generalized the result of [2] by replacing the differentiability of volatility with respect to current stock price into local Hölder continuity and improved the results of [8]. Ekström et al [7] provided a counter example of the solution not preserving convexity when the stock price follows a jump-diffusion process and studied the condition to preserve convexity. All the above mentioned works dealt with homogenous Black-Scholes (integro-) differential equations while Janson et al [12] studied preservation of convexity of solutions to general (inhomogeneous) parabolic equations under a regularity assumption of coefficients.

On the other hand, in [16] the authors provided an estimate for gradient of the solution to Black − Scholes equation with respect to stock price (which is stricter than the gradient estimate of [2], theorem 1. (i)) and applied it to comprehensive pricing of some exotic option with several expiry times. They used such a stricter estimate to prove existence of unique roots of nonlinear equations related to option prices and such roots were used as exercise prices in the next calculations. In [1] studying defaultable coupon bonds, Agliardi used the root $V^*$ of a nonlinear equation $E_N(V, T_{N-1}) = C_{N-1}$ at page 751 without any mention about the existence and uniqueness which are justified by the lemma 1 of [16, at page 253].

The purpose of this article is to study general properties of the solutions to terminal value problems of *inhomogeneous* Black-Scholes equations (with discontinuous data). We provide *representations of solutions* to terminal value problems of inhomogeneous Black-Scholes equations and study such general properties as *min-max estimates*, *gradient estimates*, *monotonicity* and *convexity* of the solutions with respect to the stock price variable. In particular, we focus on finding *representation of the gradient* (with respect to the stock price variable) of solutions to the terminal value problems with *discontinuous* terminal payoffs or inhomogeneous terms.

As Black and Sholes mentioned in [3], almost all corporate liabilities can be viewed as combinations of options. In particular we often encounter Black-Scholes equations with discontinuous maturity payoffs or inhomogeneous terms in pricing defaultable corporate bonds. See the cases with such recovery rate that $0 \leq \delta < 1$ or taxes on coupons in [1] and (4.4) in [17] when $R_e \neq R_u$ for Black-Scholes equations with discontinuous maturity payoffs. See the problems (4.11) in [17] and (4.5) in [18] for inhomogeneous Black-Scholes equations with discontinuous maturity payoffs. It is not appropriate to assume that these equations have smooth coefficients. This fact shows the necessity of studying general properties of inhomogeneous Black-Scholes equations with discontinuous maturity payoffs.

Our aim is not to establish a general theory generalizing the results of [2, 6, 8 or 11] but to develop the theory enough to cover the circumstance of inhomogeneous Black-Scholes equations with discontinuous maturity payoffs arising in the pricing problems on defaultable bonds considered in [1, 17 and 18] (or further study of this direction).

The remainder of the article is organized as follows. In section 2 we provide representations of solutions to the terminal value problems of inhomogeneous Black-Scholes equations and min-max estimates. In section 3 we study the derivatives with respect to the stock price variable in the case with *continuous* terminal payoffs and inhomogeneous terms and the solutions' *monotonicity* and *convexity*. In section 4 we study the derivatives with respect to the stock price variable in the case with *discontinuous* terminal payoffs or inhomogeneous terms, gradient estimate and the solutions' monotonicity. In section 5 we give an application to the pricing of a defaultable discrete coupon bond using our gradient estimate covering discontinuous data.

## 2. Representation of Solutions and Min-Max Estimates for Terminal Value Problems of Inhomogeneous Black-Scholes Equations

In this section we study representations of solutions to the terminal value problems of inhomogeneous Black-Scholes partial differential equations and estimate the minimum and maximum values of the solutions.

The terminal value problem of an inhomogeneous Black-Scholes partial differential equation with the risk free short rate $r(t)$, the dividend rate $q(t)$ of the underlying asset, the volatility $\sigma(t)$ of the underlying asset's price process, maturity payoff $f(x)$ and inhomogeneous term $g(x)$ is as follows:

$$\frac{\partial V}{\partial t} + \frac{\sigma^2(t)}{2} x^2 \frac{\partial^2 V}{\partial x^2} + (r(t) - q(t))x \frac{\partial V}{\partial x} - r(t)V + g(x) = 0, \quad 0 \leq t < T, \; 0 < x < \infty, \quad (2.1)$$

$$V(x, T) = f(x). \quad (2.2)$$

Throughout the whole paper, we *assume* that $r(t)$, $q(t)$, $\sigma(t)$, $f(x)$ and $g(x)$ are all *piecewise continuous* on their domains. Let denote

$$\bar{r}(t,T) = \int_t^T r(s)ds, \quad \bar{q}(t,T) = \int_t^T q(s)ds, \quad \overline{\sigma^2}(t,T) = \int_t^T \sigma^2(s)ds, \; 0 \leq t < T. \quad (2.3)$$

$$d^{\pm}(x/K, t, T) = \left(\sqrt{\overline{\sigma^2}(t,T)}\right)^{-1} \left[\ln(x/K) + \bar{r}(t,T) - \bar{q}(t,T) \pm \frac{1}{2}\overline{\sigma^2}(t,T)\right]. \quad (2.4)$$

**Theorem 1.**(Representation of Solutions to inhomogeneous Black-Scholes equations) *Assume that there exist nonnegative constants A, B, $\alpha$ and $\beta$ such that*

$$|f(x)| \leq Ax^{\alpha \ln x}, \quad |g(x)| \leq Bx^{\beta \ln x}, \; a.e. \; x > 0. \quad (2.5)$$

*Then the solution of* (2.1) *and* (2.2) *is provided as follows*:

$$V(x,t;T) = e^{-\bar{r}(t,T)} \int_0^\infty \frac{e^{-\frac{1}{2}\left[d^-\left(\frac{x}{z},t,T\right)\right]^2}}{z\sqrt{2\pi\overline{\sigma^2}(t,T)}} f(z)dz + \int_t^T e^{-\bar{r}(t,\tau)} \int_0^\infty \frac{e^{-\frac{1}{2}\left[d^-\left(\frac{x}{z},t,\tau\right)\right]^2}}{z\sqrt{2\pi\overline{\sigma^2}(t,\tau)}} g(z)dzd\tau$$

$$= x \left\{ e^{-\bar{q}(t,T)} \int_0^\infty \frac{e^{-\frac{1}{2}\left[d^+\left(\frac{x}{z},t,T\right)\right]^2}}{z^2\sqrt{2\pi\overline{\sigma^2}(t,T)}} f(z)dz + \int_t^T e^{-\bar{q}(t,\tau)} \int_0^\infty \frac{e^{-\frac{1}{2}\left[d^+\left(\frac{x}{z},t,\tau\right)\right]^2}}{z^2\sqrt{2\pi\overline{\sigma^2}(t,\tau)}} g(z)dzd\tau \right\}, \; 0 \leq t < T. \quad (2.6)$$

**Proof.** The solution of (2.1) and (2.2) is provided by the sum of the solution of the corresponding homogeneous equation of (2.1) with terminal value $f(x)$ and the solution of inhomogeneous equation with terminal value 0. By the lemma 1 of [17, at page 4] the solution of the corresponding homogeneous equation with terminal value $f(x)$ is given by the *first term* of (2.6) under the condition (2.5) on $f(x)$. Now consider the inhomogeneous equation (2.1) with terminal value 0. We can use the *Duhamel's principle* to solve it. Fix $\tau \in (0, T)$ and let $W(x,t;\tau)$ be the solution to the following terminal value problem:

$$\begin{cases} W_t + \dfrac{\sigma^2(t)}{2} x^2 W_{xx} + (r(t) - q(t))xW_x - r(t)W = 0, \; 0 \leq t < \tau, \; x > 0, \\ W(x, \tau) = g(x), \quad\quad\quad\quad\quad\quad\quad\quad\quad\quad\quad\quad\quad\quad x > 0. \end{cases}$$

This is just the homogeneous Black-Scholes equation with terminal value $g(x)$. Thus using the lemma 1 of [17] again, under the condition (2.5) on $g(x)$ its solution is given as follows:

$$W(x,t;\tau) = e^{-\bar{r}(t,\tau)} \int_0^\infty \frac{1}{\sqrt{2\pi\overline{\sigma^2}(t,\tau)}} \frac{1}{z} \exp\left\{-\frac{1}{2}\left[d^-\left(\frac{x}{z},t,\tau\right)\right]^2\right\} g(z)dz.$$

According to *Duhamel's principle*, $\int_t^T W(x,t;\tau)d\tau\ (0\leq t<T, x>0)$ is the solution to of the inhomogeneous Black-Scholes equation with terminal value 0. Thus we have (2.6).(QED)

**Remark 1**. The theorem 1 generalizes the lemma 1 of [17, at page 4] to the case of the inhomogeneous Black-Scholes equation.

**Corollary 1**. (Min-Max estimate for inhomogeneous Black-Scholes equation) *Let f(x) and g(x) be bounded and let $m(f)=\inf_x f(x), M(f)=\sup_x f(x), m(g)=\inf_x g(x), M(g)=\sup_x g(x)$. Let V(x, t) be the solution of (2.1) and (2.2). Then we have the following min-max estimates:*

i) $m(f)e^{-\bar{r}(t,T)} + m(g)\int_t^T e^{-\bar{r}(t,\tau)}d\tau \leq V(x,t) \leq M(f)e^{-\bar{r}(t,T)} + M(g)\int_t^T e^{-\bar{r}(t,\tau)}d\tau, 0\leq t\leq T.$  (2.7)

In particular if $r(t)\equiv r$ is constant, then we have

$$m(f)e^{-r(T-t)} + \frac{m(g)}{r}(1-e^{-r(T-t)}) \leq V(x,t) \leq M(f)e^{-r(T-t)} + \frac{M(g)}{r}(1-e^{-r(T-t)}), r\neq 0,$$
$$m(f) + m(g)(T-t) \leq V(x,t) \leq M(f) + M(g)(T-t),\ \ r=0,\quad 0\leq t\leq T.$$  (2.7)'

ii) *Let $E_f=\{z:f(z)>m(g)\}, F_f=\{z:f(z)<M(f)\}$. If $|E_f|>0$ or $|E_g|>0$, then we have the stricter estimate at the time before the maturity:*

$$m(f)e^{-\bar{r}(t,T)} + m(g)\int_t^T e^{-\bar{r}(t,\tau)}d\tau < V(x,t),\ 0\leq t<T.$$  (2.8)

*Similarly if $|F_f|>0$ or $|F_g|>0$, then we have*

$$\frac{\partial V}{\partial x}(x,t) < M(f)e^{-\bar{r}(t,T)} + M(g)\int_t^T e^{-\bar{r}(t,\tau)}d\tau,\ 0\leq t<T.$$  (2.9)

**Proof.** In the two improper integrals of (2.6), if we use the substitutions $y=d^-(x/z,t,T)$ and $y=d^-(x/z,t,\tau)$, respectively, then we have

$$V(x,t) = \frac{e^{-\bar{r}(t,T)}}{\sqrt{2\pi}}\int_{-\infty}^\infty e^{-y^2/2}f(xc(y,t,T))dy + \int_t^T \frac{e^{-\bar{r}(t,\tau)}}{\sqrt{2\pi}}\int_{-\infty}^\infty e^{-y^2/2}g(xc(y,t,\tau))dyd\tau.$$  (2.10)

Here

$$c(y,t,T) = \exp\left\{-y\sqrt{\overline{\sigma^2}(t,T)} + \bar{r}(t,T) - \bar{q}(t,T) - \frac{1}{2}\overline{\sigma^2}(t,T)\right\}.$$  (2.11)

From (2.10), if we consider $(\sqrt{2\pi})^{-1}\int_{-\infty}^\infty e^{-y^2/2}dy=1$, then the results are easily proved. (QED)

**Remark 2**. Maximum value estimates for Black-Scholes equations might not have been mentioned before. The reason seems that call option price itself is unbounded and such a maximum estimate is direct result of the solution representation like [13, 17]. But in the study of compound-like options such as Bermudan options or defaultable bonds with discrete default information we encounter such terminal value problems of (inhomogeneous) Black-Scholes equations that terminal values are solutions to other (inhomogeneous) Black-Scholes equations and we need to solve linear equation related to the solutions to such terminal value problems. In that case we need such a min-max estimate.

**Corollary 2**. *Let f(x) and g(x) be such functions that $f(0+), g(0+)$ and $f(+\infty), g(+\infty)$ exist. Let V(x, t) be the solution of (2.1) and (2.2). Then $V(0+, t)$ and $V(+\infty, t)$ are given as follows:*

$$V(0+, t) = f(0+)e^{-\bar{r}(t,T)} + g(0+)\int_t^T e^{-\bar{r}(t,\tau)}d\tau,$$
$$V(+\infty, t) = f(+\infty)e^{-\bar{r}(t,T)} + g(+\infty)\int_t^T e^{-\bar{r}(t,\tau)}d\tau.$$
(2.12)

**Corollary 3**. *Assume that $0 \le f(x) \le \alpha \cdot x$ and $0 \le g(x) \le \beta \cdot x$. Let V(x, t) be the solution of (2.1) and (2.2). Then we have*

$$0 \le V(x,t) \le x\left[\alpha e^{-\bar{q}(t,T)} + \beta \int_t^T e^{-\bar{q}(t,\tau)}d\tau\right], 0 \le t \le T.$$

*In particular if $q(t) \equiv q$ is constant, then we have*

$$0 \le V(x,t) \le x\left[\alpha e^{-q(T-t)} + \frac{\beta}{q}(1 - e^{-q(T-t)})\right], q \ne 0,$$
$$0 \le V(x,t) \le x[\alpha + \beta(T-t)], \quad q = 0, \quad 0 \le t \le T.$$
(2.13)

*Furthermore let $\bar{E}_h = \{x : h(x) > 0\}, \bar{F}_h = \{x : h(x) < x\}, \tilde{F}_h = \{x : h(x) < \beta \cdot x\}$. If $|\bar{E}_f| > 0$ and $|\bar{F}_f| > 0$*

*(or $|\bar{E}_g| > 0$ and $|\tilde{F}_g| > 0$), then we can replace "≤" into "<" in the above estimates.*

**Proof**. These estimates are the direct conclusion from the second formula of (2.6) and the assumption of the lemma. (QED)

## 3. The Preservation of Monotonicity and Convexity in the Case with Continuous Terminal Payoffs and Inhomogeneous Terms

In this section we study the derivatives (with respect to the stock price variable) of solutions to the terminal value problems of inhomogeneous Black-Scholes equations with *continuous* terminal payoffs and inhomogeneous terms and the solutions' monotonicity and convexity.

**Theorem 2**.(*x*-gradient of Solution of inhomogeneous Black-Scholes equation) *Let f(x) and g(x) be continuous and piecewise differentiable and satisfy*

$$|f(x), f'(x), g(x), g'(x)| \le Ax^{\alpha \ln x}, \quad a.e. \ x > 0 \ (\exists A, \alpha > 0).$$

*Let V(x, t) be the solution of (2.1) and (2.2). Then we have the following conclusions*:

i) 
$$\frac{\partial V}{\partial x}(x, t) = \frac{e^{-\bar{q}(t,T)}}{\sqrt{2\pi}} \int_{-\infty}^{\infty} e^{-\frac{1}{2}\left(y+\sqrt{\overline{\sigma^2}(t,T)}\right)^2} f'(xc(y,t,T))dy +$$

$$+ \int_t^T \frac{e^{-\bar{q}(t,\tau)}}{\sqrt{2\pi}} \int_{-\infty}^{\infty} e^{-\frac{1}{2}\left(y+\sqrt{\overline{\sigma^2}(t,\tau)}\right)^2} g'(xc(y,t,\tau))dyd\tau, \quad 0 \le t < T. \quad (3.1)$$

Here $c(y, t, T)$ is as in (2.11).

ii) If $f(x)$ and $g(x)$ are *increasing* (or *decreasing*), then $V(x, t)$ is ***x*-increasing** (or *decreasing*), *too*.

iii) If $f'(0), g'(0)$ and $f'(+\infty), g'(+\infty)$ exist, then $\partial_x V(0,t), \partial_x V(+\infty, t)$ are given as follows:

$$\frac{\partial V}{\partial x}(0, t) = f'(0)e^{-\bar{q}(t,T)} + g'(0)\int_t^T e^{-\bar{q}(t,\tau)}d\tau,$$

$$\frac{\partial V}{\partial x}(+\infty, t) = f'(+\infty)e^{-\bar{q}(t,T)} + g'(+\infty)\int_t^T e^{-\bar{q}(t,\tau)}d\tau. \quad (3.2)$$

iv) If $f'(x)$ and $g'(x)$ are bounded, then we have

$$m(f')e^{-\bar{q}(t,T)} + m(g')\int_t^T e^{-\bar{q}(t,\tau)}d\tau \le \frac{\partial V}{\partial x} \le M(f')e^{-\bar{q}(t,T)} + M(g')\int_t^T e^{-\bar{q}(t,\tau)}d\tau. \quad (3.3)$$

Furthermore if $|E_{f'}| > 0$ or $|E_{g'}| > 0$, then we have the stricter estimate on x-gradient:

$$m(f')e^{-\bar{q}(t,T)} + m(g')\int_t^T e^{-\bar{q}(t,\tau)}d\tau < \frac{\partial V}{\partial x}(x, t), \quad 0 \le t < T. \quad (3.4)$$

Similarly if $|F_{f'}| > 0$ or $|F_{g'}| > 0$, then

$$\frac{\partial V}{\partial x}(x, t) < M(f')e^{-\bar{q}(t,T)} + M(g')\int_t^T e^{-\bar{q}(t,\tau)}d\tau, \quad 0 \le t < T. \quad (3.5)$$

Here $m(f'), M(f'), E_{f'}, F_{f'}$ are same as in the corollary 1 of the theorem 1.

**Proof.** If we differentiate the both side of (2.10) on *x*, then we obtain the results of i). The results of ii), iii) and iv) are evident from (3.1). (QED)

**Remark 3**. The theorem 2, iv) is a *generalization* of the lemma 1 of [16, at page 253] which considered homogeneous Black-Sholes equations with constant coefficients. Bergman et al. [2] (theorem 1, (i)) obtained a gradient estimate for homogeneous Black-Sholes equation with stock price and time dependent coefficients but *our estimate is stricter* than it.

**Theorem 3**.(Convexity of Solutions of inhomogeneous Black-Scholes equation) *Let f(x) and g(x) be the same with those in the theorem 1 and V(x, t) the solution of* (2.1) *and* (2.2). *If both f(x) and g(x) are convex upward (or downward), then V(x, t) is convex upward (or downward) on x, too.*

**Proof.** Let assume $f(x)$ and $g(x)$ are convex upward. Then for any $\lambda \in (0, 1)$, we have

$$f(x_1 + \lambda(x_2 - x_1)) \geq f(x_1) + \lambda(f(x_2) - f(x_1)),\ g(x_1 + \lambda(x_2 - x_1)) \geq g(x_1) + \lambda(g(x_2) - g(x_1)). \quad (3.6)$$

By the formula (2.10), we have

$$V(x_1 + \lambda(x_2 - x_1), t; T) = \frac{e^{-\bar{r}(t,T)}}{\sqrt{2\pi}} \int_{-\infty}^{\infty} e^{-y^2/2} f([x_1 + \lambda(x_2 - x_1)]c(y,t,T))dy$$

$$+ \int_t^T \frac{e^{-\bar{r}(t,\tau)}}{\sqrt{2\pi}} \int_{-\infty}^{\infty} e^{-y^2/2} g([x_1 + \lambda(x_2 - x_1)]c(y,t,\tau))dy d\tau.$$

Here we use (3.6) and then use (2.10) again, then we obtain

$$V(x_1 + \lambda(x_2 - x_1), t; T) \geq V(x_1, t; T) + \lambda(V(x_2, t; T)) - V(x_1, t; T)).$$

The downward convexity is similarly proved. (QED)

**Remark 4**. The theorem 3 generalizes the well-known convexity preservation results of standard Black-Scholes equations to inhomogenous cases. Janson et al. [12] studied preservation of convexity of solutions to general inhomogeneous parabolic equations but their result *didn't cover* ours because of regularity assumption of coefficients.

## 4. Estimates of *x*-Derivatives in the Case with Discontinuous Data

In this section we study the derivatives (with respect to the stock price variable) of solutions to the terminal value problems of Black-Scholes equations with *discontinuous* terminal payoffs or inhomogeneous terms and the solutions' monotonicity. Such terminal value problems are often encountered in pricing problems of defaultable bonds with discrete default barrier, discrete default intensity and endogenous default recovery. See the cases with such recovery rate that $0 \leq \delta < 1$ or the case with taxes on coupons in [1] and (4.4) in [17, at page 17] when $R_e \neq R_u$ for homogeneous Black-Scholes equations with discontinuous terminal payoffs. See the (4.5) in [18, at page 21] and (4.11) in [17, at page 18] for inhomogeneous Black-Scholes equations with discontinuous terminal payoffs.

**Theorem 4.**(*x*-gradient of Solution of Black-Scholes equation with discontinuous maturity payoff and inhomogeneous term) *Let $f(x)$ and $g(x)$ be piecewise differentiable and satisfy*

$$|f(x), f'(x), g(x), g'(x)| \leq A x^{\alpha \ln x}, \ a.e.\ x > 0\ (\exists A, \alpha > 0).$$

*Let $V(x, t)$ be the solution of (2.1) and (2.2). Assume that $f(x)$ and $g(x)$ have the only jump discontinuities $K_f$ and $K_g$, respectively. Then we have the following conclusions*:

i) $\displaystyle \frac{\partial V}{\partial x}(x,t) = \frac{e^{-\bar{q}(t,T)}}{\sqrt{2\pi}} \left[ \int_{-\infty}^{\infty} e^{-\frac{1}{2}\left[y + \sqrt{\overline{\sigma^2}(t,T)}\right]^2} f'(xc(y,t,T))dy + \frac{e^{-\frac{1}{2}[d^+(x/K_f,t,T)]^2}}{K_f \sqrt{\overline{\sigma^2}(t,T)}} \Delta f(K_f) \right] +$

$\displaystyle + \int_t^T \frac{e^{-\bar{q}(t,\tau)}}{\sqrt{2\pi}} \left[ \int_{-\infty}^{\infty} e^{-\frac{1}{2}\left[y + \sqrt{\overline{\sigma^2}(t,\tau)}\right]^2} g'(xc(y,t,\tau))dy + \frac{e^{-\frac{1}{2}[d^+(x/K_g,t,\tau)]^2}}{K_g \sqrt{\overline{\sigma^2}(t,\tau)}} \Delta g(K_g) \right] d\tau, 0 \leq t < T.\ (4.1)$

*Here* $\Delta h(x) = h(x+0) - h(x-0)$ *is the jump of* **h** *at x and* $c(y, t, T)$ *is as in* (2.11).

ii) *If* $f(x)$ *and* $g(x)$ *are increasing* (or *decreasing*), *then* $V(x, t)$ *is **x**-increasing* (or *decreasing*), *too*.

iii) *If* $f'(0), g'(0)$ *and* $f'(+\infty), g'(+\infty)$ *exist, then* $\partial_x V(0,t), \partial_x V(+\infty, t)$ *are given as in* (3.2).

iv) *If* $f'(x)$ *and* $g'(x)$ *are bounded, then we have*

$$\partial_x V(x,t) \geq m(f') e^{-\bar{q}(t,T)} + m(g') \int_t^T e^{-\bar{q}(t,\tau)} d\tau +$$

$$+ \frac{[\Delta f(K_f)]^-}{K_f} \cdot \frac{e^{-\bar{q}(t,T)}}{\sqrt{2\pi \overline{\sigma^2}(t,T)}} + \frac{[\Delta g(K_g)]^-}{K_g} \cdot \int_t^T \frac{e^{-\bar{q}(t,\tau)}}{\sqrt{2\pi \overline{\sigma^2}(t,\tau)}} d\tau,$$

$$\partial_x V(x,t) \leq M(f') e^{-\bar{q}(t,T)} + M(g') \int_t^T e^{-\bar{q}(t,\tau)} d\tau + \frac{[\Delta f(K_f)]^+}{K_f} \cdot \frac{e^{-\bar{q}(t,T)}}{\sqrt{2\pi \overline{\sigma^2}(t,T)}}$$

$$+ \frac{[\Delta g(K_g)]^+}{K_g} \cdot \int_t^T \frac{e^{-\bar{q}(t,\tau)}}{\sqrt{2\pi \overline{\sigma^2}(t,\tau)}} d\tau, \quad 0 \leq t < T. \tag{4.2}$$

*Furthermore if* $|E_{f'}| > 0$ *or* $|E_{g'}| > 0$, *we have the stricter minimum estimate on x-gradient*

$$\partial_x V(x,t) > m(f') e^{-\bar{q}(t,T)} + m(g') \int_t^T e^{-\bar{q}(t,\tau)} d\tau +$$

$$+ \frac{[\Delta f(K_f)]^-}{K_f} \cdot \frac{e^{-\bar{q}(t,T)}}{\sqrt{2\pi \overline{\sigma^2}(t,T)}} + \frac{[\Delta g(K_g)]^-}{K_g} \cdot \int_t^T \frac{e^{-\bar{q}(t,\tau)}}{\sqrt{2\pi \overline{\sigma^2}(t,\tau)}} d\tau, \; 0 \leq t < T. \tag{4.3}$$

*Similarly if* $|F_{f'}| > 0$ *or* $|F_{g'}| > 0$, *then we have the stricter maximum estimate on x-gradient*

$$\partial_x V(x,t) < M(f') e^{-\bar{q}(t,T)} + M(g') \int_t^T e^{-\bar{q}(t,\tau)} d\tau +$$

$$+ \frac{[\Delta f(K_f)]^+}{K_f} \cdot \frac{e^{-\bar{q}(t,T)}}{\sqrt{2\pi \overline{\sigma^2}(t,T)}} + \frac{[\Delta g(K_g)]^+}{K_g} \cdot \int_t^T \frac{e^{-\bar{q}(t,\tau)}}{\sqrt{2\pi \overline{\sigma^2}(t,\tau)}} d\tau, \; 0 \leq t < T. \tag{4.4}$$

*Here* $m(f'), M(f'), E_{f'}, F_{f'}$ *are same as in the corollary 1 of the theorem 1 and* $[x]^+ = \max\{x, 0\}, [x]^- = \min\{x, 0\}$.

**Proof.** i) In (2.10), let denote $V(x,t) = U_1(x,t) + U_2(x,t)$, where

$$U_1(x,t) = \frac{e^{-\bar{r}(t,T)}}{\sqrt{2\pi}} \int_{-\infty}^{\infty} e^{-y^2/2} f(xc(y,t,T)) dy, \; U_2(x,t) = \int_t^T \frac{e^{-\bar{r}(t,\tau)}}{\sqrt{2\pi}} \int_{-\infty}^{\infty} e^{-y^2/2} g(xc(y,t,\tau)) dy d\tau.$$

Now Calculate $\partial_x U_1(x,t)$. Note that $K_f$ is the only jump discontinuity of *f(x)*. Therefore the function $h(y) = f(xc(y,t,T))$ has the only jump discontinuity $y_0$ such that $xc(y_0, t, T) = K_f$. We find $y_0$ using (2.11) and use (2.4) to get

$$y_0(x) = d^-\left(\frac{x}{K_f}, t, T\right), \quad y'_0(x) = \left(x\sqrt{\overline{\sigma^2}(t,T)}\right)^{-1}. \tag{4.5}$$

Using this $y_0$, the integral of $U_1(x,t)$ is divided into the integrals of continuous functions:

$$U_1(x,t) = \frac{e^{-\overline{r}(t,T)}}{\sqrt{2\pi}}\left[\int_{-\infty}^{y_0} e^{-y^2/2} f(xc(y,t,T))dy + \int_{y_0}^{\infty} e^{-y^2/2} f(xc(y,t,T))dy\right].$$

Using the theorem on differentiation of integral with parameter in elementary analysis, (4.5) and the fact that $c(y, t, T)$ is $y$-decreasing, we have

$$\begin{aligned}\frac{\partial U_1}{\partial x}(x,t) &= \frac{e^{-\overline{r}(t,T)}}{\sqrt{2\pi}}\left[\int_{-\infty}^{y_0} e^{-y^2/2} f'(xc(y,t,T))c(y,t,T)dy + e^{-y_0^2/2} f(xc(y_0-0,t,T))y'_0(x)\right.\\ &\quad \left.+ \int_{y_0}^{\infty} e^{-y^2/2} f'(xc(y,t,T))c(y,t,T)dy - e^{-y_0^2/2} f(xc(y_0+0,t,T))y'_0(x)\right]\\ &= \frac{e^{-\overline{q}(t,T)}}{\sqrt{2\pi}}\int_{-\infty}^{\infty} e^{-\frac{1}{2}\left(y+\sqrt{\overline{\sigma^2}(t,T)}\right)^2} f'(xc(y,t,T))dy + \frac{e^{-\overline{r}(t,T)-y_0^2/2}}{x\sqrt{2\pi\overline{\sigma^2}(t,T)}}[f(K_f+0)-f(K_f-0)].\end{aligned}$$

If we use (4.5) and (2.4), then we have

$$\frac{\partial U_1}{\partial x}(x,t) = \frac{e^{-\overline{q}(t,T)}}{\sqrt{2\pi}}\left[\int_{-\infty}^{\infty} e^{-\frac{1}{2}\left[y+\sqrt{\overline{\sigma^2}(t,T)}\right]^2} f'(xc(y,t,T))dy + \frac{\Delta f(K_f)e^{-\frac{1}{2}\left[d^+(x/K_f,t,T)\right]^2}}{K_f\sqrt{\overline{\sigma^2}(t,T)}}\right].$$

Similarly we obtain $\partial_x U_2(x,t)$:

$$\frac{\partial U_2}{\partial x}(x,t) = \int_t^T \frac{e^{-\overline{q}(t,\tau)}}{\sqrt{2\pi}}\left[\int_{-\infty}^{\infty} e^{-\frac{1}{2}\left[y+\sqrt{\overline{\sigma^2}(t,\tau)}\right]^2} g'(xc(y,t,\tau))dy + \frac{\Delta g(K_g)e^{-\frac{1}{2}\left[d^+(x/K_g,t,\tau)\right]^2}}{K_g\sqrt{\overline{\sigma^2}(t,\tau)}}\right]d\tau.$$

Adding these two equalities, we obtain (4.1).

The results of ii), iii) and iv) are evident from (4.1). (QED)

**Corollary** *Under the assumptions of theorem 4, assume that f(x) and g(x) have the finite jump discontinuities* $y_i (i=1,\cdots,n)$ *and* $z_k (k=1,\cdots,m)$, *respectively. Then we have*

$$\begin{aligned}\frac{\partial V}{\partial x}(x,t) &= \frac{e^{-\overline{q}(t,T)}}{\sqrt{2\pi}}\left[\int_{-\infty}^{\infty} e^{-\frac{1}{2}\left[y+\sqrt{\overline{\sigma^2}(t,T)}\right]^2} f'(xc(y,t,T))dy + \sum_{i=1}^{n} \frac{\Delta f(y_i)}{y_i}\cdot\frac{e^{-\frac{1}{2}\left[d^+(x/y_i,t,T)\right]^2}}{\sqrt{\overline{\sigma^2}(t,T)}}\right]\\ &+ \int_t^T \frac{e^{-\overline{q}(t,\tau)}}{\sqrt{2\pi}}\left[\int_{-\infty}^{\infty} e^{-\frac{1}{2}\left[y+\sqrt{\overline{\sigma^2}(t,\tau)}\right]^2} g'(xc(y,t,\tau))dy + \sum_{k=1}^{m} \frac{\Delta g(z_k)}{z_k}\cdot\frac{e^{-\frac{1}{2}\left[d^+(x/z_k,t,\tau)\right]^2}}{\sqrt{\overline{\sigma^2}(t,\tau)}}\right]d\tau, \ 0\le t<T.\end{aligned} \tag{4.6}$$

*Furthermore, the conclusions of* ii) *and* iii) *of theorem 4 hold and the analogues of the results of* iv) *of theorem 4 hold, too.*

**Remark 5**. The theorem 4 and its corollary generalize the theorem 2. In particular such stricter estimates as (4.3) and (4.4) (and (2.8), (2.9), (3.4), (3.5)) are important in application (see the section 3 of [16] and the following section 5). Such gradient estimates for Black-Scholes equation with discontinuous maturity payoff or inhomogeneous term can be used in calculation of default boundary in pricing real defaultable bonds with discrete coupon dates.

**Theorem 5.** (second order derivative representation of Solutions to inhomogeneous Black-Scholes equations) *Let $f(x)$ and $g(x)$ be piecewise twice differentiable and satisfy*
$$|f(x), f'(x), f''(x), g(x), g'(x), g''(x)| \leq A x^{\alpha \ln x}, \quad a.e. \ x > 0 \ (\exists A, \alpha > 0).$$
*Let $V(x, t)$ be the solution of (2.1) and (2.2). Assume that $f(x)$ and $f'(x)$ have the only jump discontinuity $K_f$ and $g(x)$ and $g'(x)$ has the only jump discontinuity $K_g$. Then we have the following second order derivative representation:*

$$\frac{\partial^2 V}{\partial x^2}(x,t;T) =$$

$$= \frac{e^{\bar{r}(t,T) - 2\bar{q}(t,T) + \overline{\sigma^2}(t,T)}}{\sqrt{2\pi}} \left\{ \int_{-\infty}^{\infty} e^{-\frac{1}{2}\left[y + 2\sqrt{\overline{\sigma^2}(t,T)}\right]^2} f''(xc(y,t,T))dy + \frac{\Delta f'(K_f)}{K_f} \cdot \frac{e^{-\frac{1}{2}\left[d^+\left(\frac{x}{K_f},t,T\right) + \sqrt{\overline{\sigma^2}(t,T)}\right]^2}}{\sqrt{\overline{\sigma^2}(t,T)}} \right.$$

$$\left. - \frac{\Delta f(K_f)}{K_f^2} \cdot \frac{1}{\overline{\sigma^2}(t,T)} e^{-\frac{1}{2}\left[d^+\left(\frac{x}{K_f},t,T\right) + \sqrt{\overline{\sigma^2}(t,T)}\right]^2} d^+\left(\frac{x}{K_f},t,T\right) \right\} +$$

$$+ \int_t^T \frac{e^{\bar{r}(t,\tau) - 2\bar{q}(t,\tau) + \overline{\sigma^2}(t,\tau)}}{\sqrt{2\pi}} \left\{ \int_{-\infty}^{\infty} e^{-\frac{1}{2}\left[y + \sqrt{\overline{\sigma^2}(t,\tau)}\right]^2} g''(xc(y,t,\tau))dy + \frac{\Delta g'(K_g)}{K_g} \cdot \frac{e^{-\frac{1}{2}\left[d^+\left(\frac{x}{K_g},t,\tau\right) + \sqrt{\overline{\sigma^2}(t,\tau)}\right]^2}}{\sqrt{\overline{\sigma^2}(t,\tau)}} \right.$$

$$\left. - \frac{\Delta g(K_g)}{K_g^2} \cdot \frac{1}{\overline{\sigma^2}(t,\tau)} e^{-\frac{1}{2}\left[d^+(x/K_g,t,\tau) + \sqrt{\overline{\sigma^2}(t,\tau)}\right]^2} d^+\left(\frac{x}{K_g},t,\tau\right) \right\} d\tau. \quad (4.7)$$

**Proof**. From the assumption and the formula (4.1), we have for $0 \leq t < T$,

$$\frac{\partial V}{\partial x}(x,t) = \frac{e^{-\bar{q}(t,T)}}{\sqrt{2\pi}} \left[ \int_{-\infty}^{\infty} e^{-\frac{1}{2}\left[y + \sqrt{\overline{\sigma^2}(t,T)}\right]^2} f'(xc(y,t,T))dy + \frac{\Delta f(K_f)}{K_f \sqrt{\overline{\sigma^2}(t,T)}} e^{-\frac{1}{2}\left[d^+(x/K_f,t,T)\right]^2} \right] +$$

$$+ \int_t^T \frac{e^{-\bar{q}(t,\tau)}}{\sqrt{2\pi}} \left[ \int_{-\infty}^{\infty} e^{-\frac{1}{2}\left[y + \sqrt{\overline{\sigma^2}(t,\tau)}\right]^2} g'(xc(y,t,\tau))dy + \frac{\Delta g(K_g)}{K_g \sqrt{\overline{\sigma^2}(t,\tau)}} e^{-\frac{1}{2}\left[d^+(x/K_g,t,\tau)\right]^2} \right] d\tau$$

$$= U_1 + U_2 + U_3 + U_4.$$

From the assumption, $h(y) = f'(xc(y,t,T))$ has the only jump discontinuity $y_0(x)$ such that $xc(y_0,t,T) = K_f$ and $H(y;\tau) = g'(xc(y,t,\tau))$ has the only jump discontinuity $y_1(x;\tau)$ such that $xc(y_1,t,\tau) = K_g$. From (2.11) and (2.4), we have

$$y_0(x) = d^-\left(\frac{x}{K_f}, t, T\right),\ y_0'(x) = \left(x\sqrt{\overline{\sigma^2}(t,T)}\right)^{-1};\ y_1(x) = d^-\left(\frac{x}{K_g}, t, \tau\right),\ y_1'(x) = \left(x\sqrt{\overline{\sigma^2}(t,\tau)}\right)^{-1}.$$

Considering this fact, differentiating $\partial_x V(x,t)$ with respect to $x$, we have

$$\frac{\partial U_1}{\partial x} = \frac{e^{-\overline{q}(t,T)}}{\sqrt{2\pi}} \frac{\partial}{\partial x} \int_{-\infty}^{\infty} e^{-\frac{1}{2}\left[y+\sqrt{\overline{\sigma^2}(t,T)}\right]^2} f'(xc(y,t,T))dy =$$

$$= \frac{e^{-\overline{q}(t,T)}}{\sqrt{2\pi}} \int_{-\infty}^{\infty} e^{-\frac{1}{2}\left[y+\sqrt{\overline{\sigma^2}(t,T)}\right]^2} f''(xc(y,t,T))c(y,t,T)dy +$$

$$+ \frac{e^{-\overline{q}(t,T)}}{\sqrt{2\pi}} e^{-\frac{1}{2}\left[y_0+\sqrt{\overline{\sigma^2}(t,T)}\right]^2} [f'(xc(y_0-0,t,T)) - f'(xc(y_0+0,t,T))] \cdot y_0'(x)$$

$$= \frac{e^{\overline{r}(t,T)-2\overline{q}(t,T)+\overline{\sigma^2}(t,T)}}{\sqrt{2\pi}} \int_{-\infty}^{\infty} e^{-\frac{1}{2}\left[y+2\sqrt{\overline{\sigma^2}(t,T)}\right]^2} f''(xc(y,t,T))dy +$$

$$+ \frac{e^{-\overline{q}(t,T)}}{x\sqrt{2\pi\overline{\sigma^2}(t,T)}} e^{-\frac{1}{2}\left[d^+(x/K_f,t,T)\right]^2} [f'(K_f+0) - f'(K_f-0)]$$

$$= \frac{e^{\overline{r}(t,T)-2\overline{q}(t,T)+\overline{\sigma^2}(t,T)}}{\sqrt{2\pi}} \left\{ \int_{-\infty}^{\infty} e^{-\frac{1}{2}\left[y+2\sqrt{\overline{\sigma^2}(t,T)}\right]^2} f''(xc(y,t,T))dy + \frac{\Delta f'(K_f)e^{-\frac{1}{2}\left[d^+\left(\frac{x}{K_f},t,T\right)+\sqrt{\overline{\sigma^2}(t,T)}\right]^2}}{K_f\sqrt{\overline{\sigma^2}(t,T)}} \right\}.$$

Similarly we have

$$\frac{\partial U_3}{\partial x} = \int_t^T \frac{e^{-\overline{q}(t,\tau)}}{\sqrt{2\pi}} \frac{\partial}{\partial x} \int_{-\infty}^{\infty} e^{-\frac{1}{2}\left[y+\sqrt{\overline{\sigma^2}(t,\tau)}\right]^2} g'(xc(y,t,\tau))dyd\tau$$

$$= \int_t^T \frac{e^{\overline{r}(t,\tau)-2\overline{q}(t,\tau)+\overline{\sigma^2}(t,\tau)}}{\sqrt{2\pi}} \left\{ \int_{-\infty}^{\infty} e^{-\frac{1}{2}\left[y+\sqrt{\overline{\sigma^2}(t,\tau)}\right]^2} g''(xc(y,t,\tau))dy + \frac{\Delta g'(K_g)e^{-\frac{1}{2}\left[d^+\left(\frac{x}{K_g},t,\tau\right)+\sqrt{\overline{\sigma^2}(t,\tau)}\right]^2}}{K_g\sqrt{\overline{\sigma^2}(t,\tau)}} \right\} d\tau.$$

On the other hand

$$\frac{\partial U_2}{\partial x} = \frac{e^{-\overline{q}(t,T)}}{\sqrt{2\pi\overline{\sigma^2}(t,T)}} \frac{\Delta f(K_f)}{K_f} \frac{\partial}{\partial x} e^{-\frac{1}{2}\left[d^+(x/K_f,t,T)\right]^2} =$$

$$= \frac{e^{-\overline{q}(t,T)}}{\sqrt{2\pi\overline{\sigma^2}(t,T)}} \frac{\Delta f(K_f)}{K_f} e^{-\frac{1}{2}\left[d^+(x/K_f,t,T)\right]^2} \left[-d^+\left(\frac{x}{K_f},t,T\right)\right] \frac{1}{x\sqrt{\overline{\sigma^2}(t,T)}}$$

$$= -\frac{e^{\overline{r}(t,T)-2\overline{q}(t,T)+\overline{\sigma^2}(t,T)}}{\overline{\sigma^2}(t,T)\sqrt{2\pi}} \frac{\Delta f(K_f)}{K_f^2} e^{-\frac{1}{2}\left[d^+(x/K_f,t,T)+\sqrt{\overline{\sigma^2}(t,T)}\right]^2} d^+\left(\frac{x}{K_f},t,T\right).$$

Similarly

$$\frac{\partial U_4}{\partial x} = \frac{\Delta g(K_g)}{K_g} \int_t^T \frac{e^{-\bar{q}(t,\tau)}}{\sqrt{2\pi\bar{\sigma^2}(t,\tau)}} \frac{\partial}{\partial x} e^{-\frac{1}{2}\left[d^+(x/K_g,t,\tau)\right]^2} d\tau$$

$$= -\frac{\Delta g(K_g)}{K_g^2} \int_t^T \frac{e^{\bar{r}(t,\tau)-2\bar{q}(t,\tau)+\bar{\sigma^2}(t,\tau)}}{\bar{\sigma^2}(t,\tau)\sqrt{2\pi}} e^{-\frac{1}{2}\left[d^+(x/K_g,t,\tau)+\sqrt{\bar{\sigma^2}(t,\tau)}\right]^2} d^+\left(\frac{x}{K_g},t,\tau\right) d\tau.$$

Thus we have proved (4.7). (QED)

## 5. Application to the Pricing of Defaultable Discrete Coupon Bonds

### 5.1 Assumptions

1) The short rate follows the *Vasicek* model

$$dr_t = (a_1 - a_2 r)dt + s_r dW_1(t) \tag{5.1}$$

under the risk neutral martingale measure and a standard Wiener process $W_1$. Under this assumption the price $Z(r, t ; T)$ of default free zero coupon bond is the solution to the problem

$$\begin{cases} \dfrac{\partial Z}{\partial t} + \dfrac{1}{2} s_r^2 \dfrac{\partial^2 Z}{\partial r^2} + (a_1 - a_2 r)\dfrac{\partial Z}{\partial r} - rZ = 0, \\ Z(r,T) = 1. \end{cases} \tag{5.2}$$

The solution is given by

$$Z(r,t;T) = e^{A(t,T) - B(t,T)r}. \tag{5.3}$$

Here $A(t, T)$ and $B(t, T)$ are given as follows [19]:

$$B(t,T) = \frac{1 - e^{-a_2(T-t)}}{a_2}, \quad A(t,T) = -\int_t^T \left[a_2 B(u,T) - \frac{1}{2}s_r^2 B^2(u,T)\right] du. \tag{5.4}$$

2) The firm value $V(t)$ follows a geometric Brown motion

$$dV(t) = (r_t - b)V(t)dt + s_V V(t) dW_2(t)$$

under the risk neutral martingale measure and a standard Wiener process $W_2$ and $E(dW_1, dW_2) = \rho dt$. The firm continuously pays out dividend in rate $b \geq 0$ (constant) for a unit of firm value.

3) A firm issues a *corporate bond* (debt) with maturity **T** and face value **F** (unit of currency). $0 = T_0 < T_1 < \cdots < T_{N-1} < T_N = T$ and $T_i$ ($i = 1,\ldots, N$) are the discrete coupon dates. At time $T_i$ ($i = 1,\ldots, N-1$) the bond holder receives the coupon of quantity $C_i \cdot Z(r,T_i;T)$ (unit of currency) from the firm, at time $T_N = T$ the bond holder receives the face value **F** and the last coupon $C_N$ (unit of currency) and $C_k \geq 0$. This means that the time *T*-value of the *sum of the face value and the coupons* of the bond is $F + \Sigma_{k=1}^N C_k$.

4) The *expected default* occurs only at time $T_i$ ($i =1,\ldots, N$) when the firm value is not

enough to pay debt and coupon. If the expected default occurs, the bond holder receives $\delta \cdot V$ as *default recovery*. Here $\delta$ is called a *fractional recovery rate* of firm value at default.

5) The unexpected default can occur at any time. The unexpected default probability in the time interval $[t, t+\Delta t]$ is $\lambda \Delta t$. Here the *default intensity* $\lambda \geq 0$ is a constant. If the unexpected default occurs at time $t \in (T_i, T_{i+1})$, the bond holder receives $\min\{\delta \cdot V_t, (F + \Sigma_{k=i+1}^{N} C_k) \cdot Z(r, t; T)\}$ as default recovery. (Here the *reason* why the expected default recovery and unexpected recovery are *given* in *different* forms is to avoid the possibility of *paying* at time $t \in (T_i, T_{i+1})$ *more than the current price of risk free zero coupon bond* with the face value $F + \Sigma_{k=i+1}^{N} C_k$ as a default recovery when the unexpected default event occurs. See [17] and [18].)

6) The price of our corporate bond is given by a sufficiently smooth function $B_i(V, r, t)$ in the subinterval $(T_i, T_{i+1})$, ($i = 0, \cdots, N-1$).

### 5.2 The Pricing Model for Defaultable Discrete Coupon Bond

According to [19], under the above assumptions 1) and 2), the price **B** of defaultable bond with a constant default intensity $\lambda$ and unexpected default recovery $R_{ud}$ satisfies the following PDE:

$$\frac{\partial B}{\partial t} + \frac{1}{2}\left[s_V^2 V^2 \frac{\partial^2 B}{\partial V^2} + 2\rho\, s_V s_r V \frac{\partial^2 B}{\partial V \partial r} + s_r^2 \frac{\partial^2 B}{\partial r^2}\right] + (r-b)V \frac{\partial B}{\partial V} + a_r \frac{\partial B}{\partial r} - (r+\lambda)B + \lambda R_{ud} = 0.$$

Therefore under the above assumptions 3), 5) and 6) we can know that our bond price $B_i(V, r, t)$ ($i = 0, \cdots, N-1$) satisfies the following PDE in the subinterval $(T_i, T_{i+1})$:

$$\frac{\partial B_i}{\partial t} + \frac{1}{2}\left[s_V^2 V^2 \frac{\partial^2 B_i}{\partial V^2} + 2\rho s_V s_r V \frac{\partial^2 B_i}{\partial V \partial r} + s_r^2 \frac{\partial^2 B_i}{\partial r^2}\right] + (r-b)V \frac{\partial B_i}{\partial V} + (a_1 - a_2 r)\frac{\partial B_i}{\partial r} \quad (5.5)$$
$$- (r+\lambda)B_i + \lambda \min\left\{\delta \cdot V, (F + \Sigma_{k=i+1}^{N} C_k)Z(r, t; T)\right\} = 0.$$

Now let derive the terminal value conditions from the conditions 3) and 4). At time $T = T_N$, the expected default occurs and the bond holder receives $\delta \cdot V_T$ if $V_T < F + C_N$. Otherwise the bond holder receives $F + C_N$. Thus we have

$$B_{N-1}(V, r, T_N) = (F + C_N) \cdot 1\{V \geq F + C_N\} + \delta V \cdot 1\{V < F + C_N\}. \quad (5.6)$$

When $i \in \{1, \ldots, N-1\}$, assume that the bond price $B_i(V, r, t)$ in the subinterval $(T_i, T_{i+1}]$ are already calculated. The $B_i(V, r, t)$ is just the price of our bond (the quantity of debt) in the subinterval $(T_i, T_{i+1}]$ under the condition that the default didn't occur before or at the time $T_i$. So the fact that the expected default did not occur at time $T_i$ means that firm value must *not be less* than $B_i(V, r, T_i)$ after paying coupon at time $T_i$, that is, $V_{T_i} \geq B_i(V, r, T_i) + C_i Z(r, T_i; T)$. Thus we have

$$B_i(V, r, T_{i+1}) = [B_{i+1}(V, r, T_{i+1}) + C_{i+1} Z(r, T_{i+1}; T)] 1\{V > B_{i+1}(V, r, T_{i+1}) + C_{i+1} Z(r, T_{i+1}; T)\} +$$
$$+ \delta V \cdot 1\{V \leq B_{i+1}(V, r, T_{i+1}) + C_{i+1} Z(r, T_{i+1}; T)\}, i = 0, \ldots, N-2. \quad (5.7)$$

The problem (5.5), (5.6), (5.7) is just the pricing model of our defaultable discrete coupon bond.

**Remark 6**. 1) As shown in the above in our model the consideration of *unexpected default risk* and dividend of firm value is added to the model on defaultable discrete coupon bond of [1]. Another difference from [1]'s approach is that the bond is considered as a derivative of risk free rate $r$ (but not a derivative of default free zero coupon bond) and the coupon prior to the maturity is provided as a discounted value of the maturity (predetermined) value.

2) The equation (5.5) is just the same type with (3.7) in [17]. In (5.6) the *default barrier is explicitly shown*. But in (5.7) the default conditions don't show the default barrier explicitly. So in this stage, it is still not clear to find the similarity of this problem with the problem of [17] but through more careful consideration we can use the method of [17] to get the pricing formula of our defaultable discrete coupon bond (see the following subsection 5.3). As shown in bellow, solving this problem is an implementation of the method of [17].

We introduce the following notation for simplicity of pricing formulae.

$$\bar{c}_N = F + C_N; \quad \bar{c}_i = C_i, i = 1, \cdots, N-1;$$
$$\Phi_i = F + \Sigma_{k=i+1}^N C_k; \qquad K_N = F + C_N; \qquad (5.8)$$
$$\Delta T_i = T_{i+1} - T_i, i = 0, \cdots, N-1.$$

That is, $\bar{c}_i$ is the time $T_N$-value of the payoff to bondholders at time $T_i$ ($i = 1, \ldots, N$) and $\Phi$ is the time $T_N$-value of the sum of the face value and all coupons of the bond. $K_N$ denotes the default barrier at time $T_N$ as in [17].

### 5.3 The Pricing Formulae of the Defaultable Discrete Coupon Bond

**Theorem 6.** *The solution of* (5.5), (5.6) *and* (5.7) *is provided as follows:*

$$B_i(V, r, t) = Z(r, t; T_N) \cdot u_i(V / Z(r, t; T_N), t), T_i < t \leq T_{i+1}, \quad i = 0, \cdots, N-1. \qquad (5.9)$$

*Here* $u_i(x, t), T_i < t \leq T_{i+1}, \quad i = 0, \cdots, N-1$ *are the solutions to the following problems*

$$\frac{\partial u_i}{\partial t} + \frac{1}{2}\sigma^2(t) x^2 \frac{\partial^2 u_i}{\partial x^2} - bx\frac{\partial u_i}{\partial x} - \lambda u_i + \lambda \min\{\Phi_i, \delta x\} = 0, \; x > 0, \; T_i < t < T_{i+1}, i = 0, \cdots, N-1,$$

$$u_{N-1}(x, T_N) = \bar{c}_N \cdot 1\{x > K_N\} + \delta x \cdot 1\{x \leq K_N\},$$

$$u_i(x, T_{i+1}) = [u_{i+1}(x, T_{i+1}) + \bar{c}_{i+1}] \cdot 1\{x > u_{i+1}(x, T_{i+1}) + \bar{c}_{i+1}\} + \delta x \cdot 1\{x \leq u_{i+1}(x, T_{i+1}) + \bar{c}_{i+1}\}, i = \overline{0, N-2}.$$

Here $\sigma(t)$ is given as follows when $B(t, T)$ is as (5.4):

$$\sigma(t) = \sqrt{s_V^2 + 2\rho s_V s_r \cdot B(t, T) + s_r^2 B^2(t, T)} > 0. \qquad (5.10)$$

$u_i(x, t), T_i < t \leq T_{i+1}, i = 0, \cdots, N-1$ *are provided as combinations of higher order binaries and their integrals.*

(i) *If every nonlinear equation*

$$x = u_i(x, T_i) + \bar{c}_i, \; x \geq 0 \quad (i = 1, \cdots, N-1) \qquad (5.11)$$

*has unique root* $K_i (i = 1, \cdots, N-1)$, *respectively, then* $u_i(x, t), \; i = 0, \cdots, N-1$ *are provided as follows:*

$$u_i(x, t) = \sum_{m=i}^{N-1} \left[ \bar{c}_{m+1} B_{K_{i+1} \cdots K_{m+1}}^{+ \; \cdots \; +}(x, t; T_{i+1}, \cdots, T_{m+1}) + \delta \cdot A_{K_{i+1} \cdots K_m K_{m+1}}^{+ \; \cdots \; + \; -}(x, t; T_{i+1}, \cdots, T_m, T_{m+1}) \right]$$

$$+ \lambda \sum_{m=i+1}^{N-1} \int_{T_m}^{T_{m+1}} \left[ \Phi_m \cdot B^{+\cdots++}_{K_{i+1}\cdots K_m \frac{\Phi_m}{\delta}}(x,t;T_{i+1},\cdots,T_m,\tau) + \delta \cdot A^{+\cdots+-}_{K_{i+1}\cdots K_m \frac{\Phi_m}{\delta}}(x,t;T_{i+1},\cdots,T_m,\tau) \right] d\tau$$

$$+ \lambda \int_{t}^{T_{i+1}} \left[ \Phi_i \cdot B^{+}_{\Phi_i/\delta}(x,t;\tau) + \delta \cdot A^{-}_{\Phi_i/\delta}(x,t;\tau) \right] d\tau, \quad T_i < t \leq T_{i+1}, \; x > 0. \tag{5.12}$$

Here $B^{+\cdots+}_{K_1\cdots K_m}(x,t;T_1,\cdots,T_m)$ and $A^{+\cdots+-}_{K_1\cdots K_{m-1}K_m}(x,t;T_1,\cdots,T_{m-1},T_m)$ are the price of m-th order bond and asset binaries with risk free rate $\lambda$, dividend rate $\lambda+b$ and volatility $\sigma(t)$ (the theorem 1 of [17, pages 5~6]). *In particular the initial price of the bond is given by*

$$B_0 = B_0(V_0, r_0, 0) =$$
$$= Z_0 \left\{ \sum_{m=0}^{N-1} \left[ \overline{c}_{m+1} B^{+\cdots+}_{K_1\cdots K_{m+1}}(V_0/Z_0, 0; T_1,\cdots,T_{m+1}) + \delta \cdot A^{+\cdots+-}_{K_1\cdots K_m K_{m+1}}(V_0/Z_0, 0; T_1,\cdots,T_m, T_{m+1}) \right. \right.$$
$$\left. \left. + \lambda_m \int_{T_m}^{T_{m+1}} [\Phi_m \cdot B^{+\cdots++}_{K_1\cdots K_m \Phi_m/\delta}(V_0/Z_0, 0; T_1,\cdots,T_m,\tau) + \delta \cdot A^{+\cdots+-}_{K_1\cdots K_m \Phi_m/\delta}(V_0/Z_0, 0; T_1,\cdots,T_m,\tau)] d\tau \right] \right\}.$$

(5.13)

Here $Z_0 = Z(r_0, 0; T)$.

(ii) *If some of the equations* $x = u_{i+1}(x, T_{i+1}) + \overline{c}_{i+1}$ ($i \in \{0,\cdots,N-2\}$) *have several roots* $K_{i+1}(j)$, $j=1, \ldots, M_i$, *then the corresponding* $u_i(x,t)$ *is provided as combinations of higher order binaries* (*related to* $K_{i+1}(j), j=1, \ldots, M$) *and their integrals*.

**Remark 7**. Generally, the nonlinear equations (5.11) have an odd number of roots. See the proof. For example In the case when $\delta = 1$, all the nonlinear equations (5.11) have the unique roots $K_i$, respectively. When $0 \leq \delta < 1$, if the volatility $\sigma(t)$ of the *relative* price of the firm value $x = V/Z$ is enough large, that is, if there exist a sequence $\delta = d_N < d_{N-1} < \cdots < d_1 < 1$ such that

$$\sqrt{\overline{\sigma^2}(T_k, T_{k+1})} \geq (1-\delta)/\left[\sqrt{2\pi}(d_k - d_{k+1})\right]^{-1} \text{ if } \lambda = b = 0 \; (k=1,\cdots,N-1),$$

$$\sqrt{\overline{\sigma^2}(T_k, T_{k+1})} \geq \frac{(1-\delta)e^{-(\lambda+b)\Delta T_k}}{\sqrt{2\pi}\left\{d_k - \left[d_{k+1}e^{-(\lambda+b)\Delta T_k} + \frac{\lambda\delta}{\lambda+b}\left(1 - e^{-(\lambda+b)\Delta T_k}\right)\right]\right\}} \text{ if } \lambda + b > 0, \tag{5.14}$$

then all the nonlinear equations (5.11) have the unique roots $K_i$, respectively, too. In the case when (5.14) do not satisfy, if the coupons at time $T_i$ ($i=1,\ldots,N-1$) is not too large, then all the nonlinear equations (5.11) have the unique roots $K_i$. (See Lemmas 1 and 2 in the proof.) The case when some of the equations (5.11) have several roots seems not appropriate in financial meaning. (See the remark 9 in the proof.)

**Proof**: Now we solve the problem (5.5), (5.6), (5.7). For $i=N-1$, the problem (5.5) and (5.6) becomes as follows:

$$\frac{\partial B_{N-1}}{\partial t} + \frac{1}{2}\left[\sigma^2(t)V^2 \frac{\partial^2 B_{N-1}}{\partial V^2} + 2\rho s_V(t)s_r(t)V \frac{\partial^2 B_{N-1}}{\partial V \partial r} + s_r^2(t)\frac{\partial^2 B_{N-1}}{\partial r^2}\right] + (r-b)V\frac{\partial B_{N-1}}{\partial V}$$

$$+ (a_1 - a_2 r)\frac{\partial B_{N-1}}{\partial r} - (r+\lambda)B_{N-1} + \lambda \min\{\delta \cdot V, \Phi_{N-1} \cdot Z(r,t;T)\} = 0, \ T_{N-1} < t < T_N, x > 0$$

$$B_{N-1}(V, r, T_N) = \bar{c}_N \cdot 1\{V \geq K_N\} + \delta V \cdot 1\{V < K_N\}, \quad x > 0.$$

Here if we let $i = N-1$ and use the *change of numeraire*

$$x = V/Z(r,t;T), \quad u_i(x,t) = B_i(V, r, t)/Z(r,t;T), \ T_i < t \leq T_{i+1} \tag{5.15}$$

and consider (5.2), then we have

$$\begin{cases} \dfrac{\partial u_{N-1}}{\partial t} + \dfrac{1}{2}\sigma^2(t)x^2 \dfrac{\partial^2 u_{N-1}}{\partial x^2} - bx\dfrac{\partial u_{N-1}}{\partial x} - \lambda u_{N-1} + \lambda \min\{\Phi_{N-1}, \delta x\} = 0, T_{N-1} < t < T_N, \ x > 0, \\ u_{N-1}(x, T_N) = \bar{c}_N \cdot 1\{x > K_N\} + \delta x \cdot 1\{x \leq K_N\}, \quad x > 0. \end{cases}$$

This problem is just the same problem with the (2.1) and (2.2) with the coefficients $r(t) \equiv \lambda, q(t) \equiv \lambda + b, \sigma(t)$ and the inhomogeneous term

$$g(x) = \lambda \min\{\Phi_{N-1}, \delta x\} = \lambda(\Phi_{N-1} \cdot 1\{x > \Phi_{N-1}/\delta\} + \delta \cdot x \cdot 1\{x < \Phi_{N-1}/\delta\}).$$

Furthermore the terminal value

$$f(x) = \bar{c}_N \cdot 1\{x > K_N\} + \delta x \cdot 1\{x \leq K_N\}$$

is a discontinuous function if $\delta < 1$.

Using the theorem 1, we have

$$u_{N-1}(x,t;T_N) = e^{-\lambda(T_N-t)} \int_0^\infty \frac{e^{-\frac{\left(\ln\frac{x}{z} - b(T_N-t) - \frac{1}{2}\overline{\sigma^2}(t,T_N)\right)^2}{2\overline{\sigma^2}(t,T_N)}}}{z\sqrt{2\pi\overline{\sigma^2}(t,T_N)}} (\bar{c}_N \cdot 1\{z > K_N\} + \delta \cdot z \cdot 1\{z \leq K_N\})dz$$

$$+ \lambda \int_t^{T_N} e^{-\lambda(\tau-t)} \int_0^\infty \frac{e^{-\frac{\left(\ln\frac{x}{z} - b(\tau-t) - \frac{1}{2}\overline{\sigma^2}(t,\tau)\right)^2}{2\overline{\sigma^2}(t,\tau)}}}{z\sqrt{2\pi\overline{\sigma^2}(t,\tau)}} (\Phi_{N-1} \cdot 1\{z > \Phi_{N-1}/\delta\} + \delta \cdot z \cdot 1\{z < \Phi_{N-1}/\delta\})dzd\tau.$$

From the definition of bond and asset binaries [4, 16, 17] we can know that the above two terms are just the representation of linear combinations of bond and asset binaries and the integrals of them. Thus we have

$$u_{N-1}(x,t) = \bar{c}_N B^+_{K_N}(x,t;T_N) + \delta A^-_{K_N}(x,t;T_N)$$

$$+ \lambda \int_t^{T_N} [\Phi_{N-1} \cdot B^+_{\Phi_{N-1}/\delta}(x,t;\tau) + \delta A^-_{\Phi_{N-1}/\delta}(x,t;\tau)]d\tau, \ T_{N-1} < t \leq T_N, \ x > 0. \tag{5.16}$$

Here $B^+_K(x,t;T)$ and $A^-_K(x,t;T)$ are the prices of bond and asset binary options with the coefficients $r(t) \equiv \lambda, q(t) \equiv \lambda + b, \sigma(t)$ and maturity $T$, which are provided by the theorem 1 of [17, at page 5~6]. Returning to original variables using (5.15), we obtain $B_{N-1}$:

$$B_{N-1}(V, r, t) = Z(r,t;T) \cdot u_{N-1}(V/Z(r,t;T), t), \ (T_{N-1} < t \leq T_N, \ x > 0).$$

For the next step of study we investigate $u_{N-1}(x, T_{N-1})$ and $\partial_x u_{N-1}(x, T_{N-1})$. Note that $m(f) = m(g) = 0, M(f) = \bar{c}_N = K_N, M(g) = \Phi_{N-1}$, $f$ and $g$ are increasing, $f(x) \leq x$ and $g(x) \leq$ $\lambda\delta\cdot x$. From the corollary 1 and corollary 3 of the theorem 1, ii) of the theorem 4, we know that $u_{N-1}(x, T_{N-1})$ is bounded and increasing. Furthermore we have

$$u_{N-1}(0, T_{N-1}) = 0 < u_{N-1}(x, T_{N-1}) < \min\{u_{N-1}(+\infty, T_{N-1}), x\}. \qquad (5.17)$$

Here $u_{N-1}(+\infty, T_{N-1}) = e^{-\lambda\Delta T_{N-1}} K_N + (1 - e^{-\lambda\Delta T_{N-1}})\Phi_{N-1}$. (See the figure 1.)

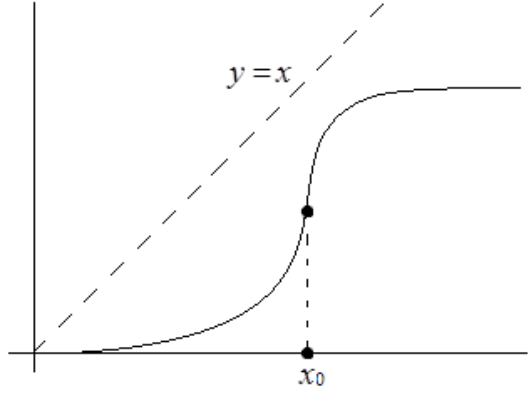

Figure 1. Convex and concave intervals of $u_{N-1}(x, T_{N-1})$ that is bellow $y = x$.

$f(x)$ has the jump $\Delta f(K_N) = (1-\delta)K_N$ at $K_N$ and $m(f') = m(g') = 0$, $M(f') = \delta = d_N$, $M(g') = \lambda\delta$. Thus using iv) of the theorem 4, we have

$$0 < \partial_x u_{N-1}(x, T_{N-1}) < d_N e^{-(\lambda+b)\Delta T_{N-1}} + \frac{\lambda\delta}{\lambda+b}(1 - e^{-(\lambda+b)\Delta T_{N-1}}) + \frac{(1-\delta)e^{-(\lambda+b)\Delta T_{N-1}}}{\sqrt{2\pi\overline{\sigma^2}(T_{N-1}, T_N)}} \quad \text{if } \lambda+b > 0,$$

$$0 < \partial_x u_{N-1}(x, T_{N-1}) < d_N + \frac{1-\delta}{\sqrt{2\pi\overline{\sigma^2}(T_{N-1}, T_N)}} \quad \text{if } \lambda = b = 0. \qquad (5.18)$$

Therefore if $\delta = 1$, then

$$0 < \partial_x u_{N-1}(x, T_{N-1}) < 1. \qquad (5.19)$$

If $\delta < 1$ and the condition (5.14) with $k = N-1$ holds, then we have

$$0 < \partial_x u_{N-1}(x, T_{N-1}) < d_{N-1} < 1 \qquad (5.20)$$

That is, (5.14) is a sufficient condition for (5.19) to be true.

Now let study the convexity of $u_{N-1}(x, T_{N-1})$. Note that $f'$ has jump $-\delta$ at $x = K_N$, $g'$ has jump $-\lambda_{N-1}\delta$ at $x = \Phi_{N-1}/\delta$ and $f''(x) = g''(x) = 0$, a.e. $x > 0$. Using the theorem 5, we have

$$\frac{\partial^2 u_{N-1}}{\partial x^2}(x, T_{N-1}) = \frac{-\delta \cdot e^{-(\lambda+2b)\Delta T_{N-1} + \overline{\sigma^2}(T_{N-1}, T_N)}}{K_N\sqrt{2\pi\overline{\sigma^2}(T_{N-1}, T_N)}} e^{-\frac{1}{2}\left[d^+\left(\frac{x}{K_N}, T_{N-1}, T_N\right) + \sqrt{\overline{\sigma^2}(T_{N-1}, T_N)}\right]^2}$$

$$- \frac{\lambda\delta^2}{\Phi_{N-1}} \int_t^T \frac{e^{-(+2b)(\tau - T_{N-1}) + \overline{\sigma^2}(T_{N-1}, \tau)}}{\sqrt{2\pi\overline{\sigma^2}(T_{N-1}, \tau)}} e^{-\frac{1}{2}\left[d^+\left(\frac{\delta x}{\Phi_{N-1}}, T_{N-1}, \tau\right) + \sqrt{\overline{\sigma^2}(T_{N-1}, \tau)}\right]^2} d\tau$$

$$- \frac{(1-\delta)e^{-(\lambda+2b)\Delta T_{N-1} + \overline{\sigma^2}(T_{N-1}, T_N)}}{\sqrt{2\pi}K_N\overline{\sigma^2}(T_{N-1}, T_N)} d^+\left(\frac{x}{K_N}, T_{N-1}, T_N\right) e^{-\frac{1}{2}\left[d^+\left(\frac{x}{K_N}, T_{N-1}, T_N\right) + \sqrt{\overline{\sigma^2}(T_{N-1}, T_N)}\right]^2}.$$

From this, we can know that if δ=1 then $\partial_x^2 u_{N-1}(x,T_{N-1}) < 0$, that is, $u_{N-1}(x,T_{N-1})$ is x-upward convex and if $0 \leq \delta < 1$ then there exists unique point $x_0$ such that

$$\partial_x^2 u_{N-1}(x_0, T_{N-1}) = 0 \ ; \ \partial_x^2 u_{N-1}(x, T_{N-1}) > 0, x < x_0; \ \partial_x^2 u_{N-1}(x, T_{N-1}) < 0, x > x_0.$$

(See the figure 2.) That is, if $x < x_0$, then $u_{N-1}(x, T_{N-1})$ is downward convex and if $x > x_0$, then $u_{N-1}(x, T_{N-1})$ is upward convex. (See the figure 1.)

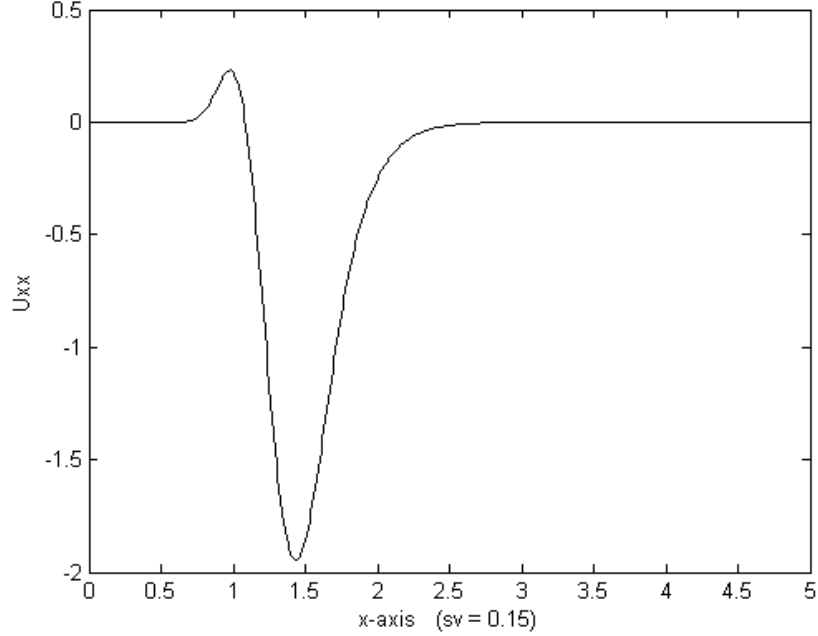

Figure 2. A graph of $\partial_x^2 u_{N-1}(x, T_{N-1})$, $\delta = 0.8$, $\lambda = 0.1$, $b = 0.05$, $T_{N-1} = 2$, $T_N = 3$, $K_N = 1.3$, $\Phi_{N-1} = 1.6$, $a_2 = 0.379$, $s_r = 0.077$, $\rho = 0.5$, $s_V = 0.15$

Now since we have enough information about $u_{N-1}(x, T_{N-1})$, let consider the case when $i = N-2$ in the problem (5.5) and (5.7).

$$\frac{\partial B_{N-2}}{\partial t} + \frac{1}{2}\left[s_V^2(t)V^2 \frac{\partial^2 B_{N-2}}{\partial V^2} + 2\rho s_V(t)s_r(t)V \frac{\partial^2 B_{N-2}}{\partial V \partial r} + s_r^2(t) \frac{\partial^2 B_{N-2}}{\partial r^2}\right] + (r-b)V \frac{\partial B_{N-2}}{\partial V}$$

$$+ a_r(r,t)\frac{\partial B_{N-2}}{\partial r} - (r + \lambda_{N-2})B_{N-2} + \lambda_{N-2} \min\{\delta \cdot V, \Phi_{N-2} \cdot Z(r,t;T)\} = 0, \ T_{N-2} < t < T_{N-1},$$

$$B_{N-2}(V, r, T_{N-1}) =$$
$$= [B_{N-1}(V, r, T_{N-1}) + \bar{c}_{N-1}Z(r, T_{N-1}, T_N)] \cdot 1\{V > B_{N-1}(V, r, T_{N-1}) + \bar{c}_{N-1}Z(r, T_{N-1}, T_N)\}$$
$$+ \delta V \cdot 1\{V \leq B_{N-1}(V, r, T_{N-1}) + \bar{c}_{N-1}Z(r, T_{N-1}, T_N)\}.$$

In this problem, we use (5.8) with $i=N-2$ to get

$$\frac{\partial u_{N-2}}{\partial t} + \frac{1}{2}S_X^2(t)x^2 \frac{\partial^2 u_{N-2}}{\partial x^2} - bx\frac{\partial u_{N-2}}{\partial x} - \lambda u_{N-2} + \lambda \min\{\Phi_{N-2}, \delta x\} = 0, \quad (5.21)$$

$$u_{N-2}(x, T_{N-1}) = [u_{N-1}(x, T_{N-1}) + \bar{c}_{N-1}] \cdot 1\{x > u_{N-1}(x, T_{N-1}) + \bar{c}_{N-1}\} + \delta x \cdot 1\{x \leq u_{N-1}(x, T_{N-1}) + \bar{c}_{N-1}\}. \quad (5.22)$$

The equation (5.21) is just the same type with the Black-Scholes equation for $u_{N-1}$ but the *inhomogenous term* is

$$g(x) = \lambda \min\{\Phi_{N-2}, \delta x\}$$

and the terminal condition (5.22) is *not a standard* binary type. To make it into a standard binary type, we now consider the equation $x = u_{N-1}(x, T_{N-1}) + \bar{c}_{N-1}$. If (5.19) holds, then the equation $x = u_{N-1}(x, T_{N-1}) + \bar{c}_{N-1}$ has unique root $K_{N-1} \geq 0$ and we have $x > u_{N-1}(x, T_{N-1}) + \bar{c}_{N-1} \Leftrightarrow x > K_{N-1}$. Note that $\bar{c}_{N-1} = 0 \Rightarrow K_{N-1} = 0$. If (5.19) does not hold, then there exists a $x_1$ such that $\partial_x u_{N-1}(x_1, T_{N-1}) \geq 1$ and we can know that the equation $x = u_{N-1}(x, T_{N-1}) + \bar{c}_{N-1}$ can have one root $K_{N-1}$ or three roots $K_{N-1}(1), K_{N-1}(2)$ and $K_{N-1}(3)$ such that $K_{N-1}(1) < K_{N-1}(2) < K_{N-1}(3)$. (See the figure 3.)

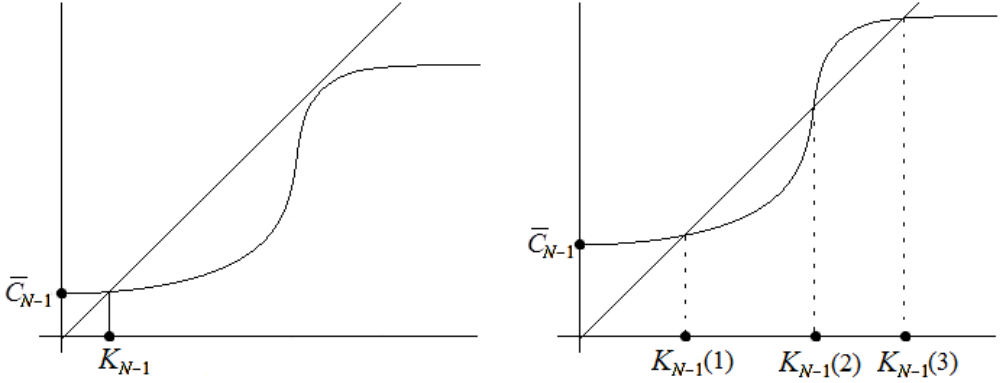

Figure 3. roots of $x = u_{N-1}(x, T_{N-1}) + \bar{c}_{N-1}$.

The figure 3 allows us to provide a *necessary and sufficient condition* for the equation $x = u_{N-1}(x, T_{N-1}) + \bar{c}_{N-1}$ to have unique root.

**Lemma 1**. *The equation* $x = u_{N-1}(x, T_{N-1}) + \bar{c}_{N-1}$ *has unique root if and only if one of the three conditions satisfies.*

(i) *If* $x_1$ *is the maximum root of* $\partial_x u_{N-1}(x_1, T_{N-1}) = 1$, *then*
$$\bar{c}_{N-1} < x_1 - u_{N-1}(x_1, T_{N-1}). \tag{5.23}$$

(ii) *If* $x_0$ *is the minimum root of* $\partial_x u_{N-1}(x_0, T_{N-1}) = 1$, *then*
$$\bar{c}_{N-1} > x_0 - u_{N-1}(x_0, T_{N-1}). \tag{5.24}$$

**Remark 8**. The lemma 1 provides a *theoretical* necessary and sufficient condition for the equation $x = u_{N-1}(x, T_{N-1}) + \bar{c}_{N-1}$ to have unique root. For example (5.24) seems *not appropriate* in financial meaning. But (5.23) seems to provide an *appropriate* (in financial meaning) *upper bound* for the *coupon*.

If the $x = u_{N-1}(x, T_{N-1}) + \bar{c}_{N-1}$ has one root $K_{N-1}$ then the terminal value (5.22) can be written as follows:
$$u_{N-2}(x, T_{N-1}) = [u_{N-1}(x, T_{N-1}) + \bar{c}_{N-1}] \cdot 1\{x > K_{N-1}\} + \delta x \cdot 1\{x \leq K_{N-1}\} = f(x). \tag{5.25}$$

If the equation $x = u_{N-1}(x, T_{N-1}) + \bar{c}_{N-1}$ have three roots $K_{N-1}(1) < K_{N-1}(2) < K_{N-1}(3)$, then the terminal value (5.22) can be written as follows (See the figure 4.):
$$\begin{aligned} u_{N-2}(x, T_{N-1}) &= [u_{N-1}(x, T_{N-1}) + \bar{c}_{N-1}] \cdot [1\{K_{N-1}(1) \leq x < K_{N-1}(2)\} + 1\{x \geq K_{N-1}(3)\}] + \\ &+ \delta x \cdot [1\{x \leq K_{N-1}(1)\} + 1\{K_{N-1}(2) \leq x < K_{N-1}(3)\}] = f(x). \end{aligned} \tag{5.26}$$

The problem (5.21) with (5.25) or (5.26) is just same type with the problem for $u_{N-1}$. Its terminal value *f(x)* is just the binary type. So we can get the solution-representation of the problem by the method of [17].

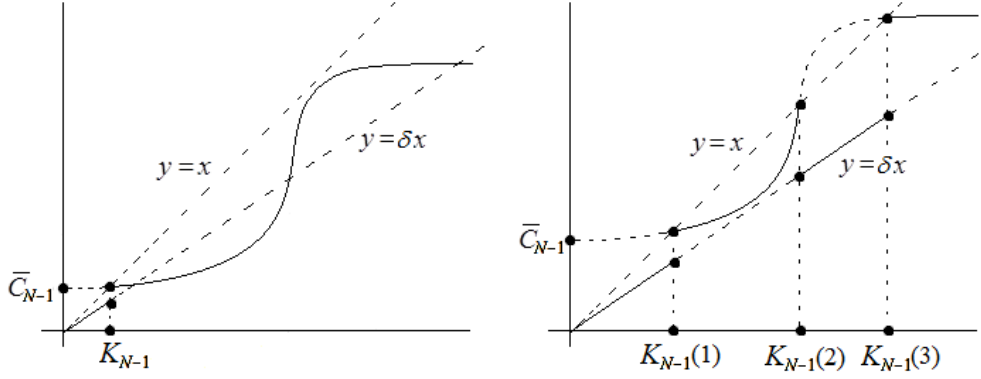

Figure 4.  $u_{N-2}(x, T_{N-1}) = f(x)$

**Remark 9**. As shown in (5.25) and (5.26), $K_{N-1}$ or $K_{N-1}(1), K_{N-1}(2)$ and $K_{N-1}(3)$ give the *default barrier* at time $T_{N-1}$. In (5.26), $K_{N-1}(1) \leq x < K_{N-1}(2)$ is a survival interval but $K_{N-1}(2) \leq x < K_{N-1}(3)$ is a default interval for $x = V/Z$. It seems *not compatible* with financial reality. Thus we *will not consider* (5.26) in what follows.

Now we postpone finding the representation of $u_{N-2}(x,t)$ for a while and first estimate $u_{N-2}$ and $\partial_x u_{N-2}$. From (5.25) and (5.17), we have $m(f) = 0$, $M(f) = e^{-\lambda \Delta T_{N-1}} K_N + (1 - e^{-\lambda \Delta T_{N-1}}) \Phi + \bar{c}_{N-1}$, *f is* increasing and $f(x) \leq x$. Using again the corollary 1 and corollary 3 of the theorem 1, ii) of the theorem 4, $u_{N-2}(x, T_{N-2})$ is bounded, increasing and $0 < u_{N-2}(x, T_{N-2}) < x$. (***Note*** that when *f(x)* is given by (5.26), *f(x)* is not increasing and thus $u_{N-2}(x, T_{N-2})$ is not increasing either but other all conclusions on $u_{N-2}(x, T_{N-2})$ still hold.)

On the other hand, from (5.25), *f(x)* has the jump $\Delta f(K_{N-1}) = (1 - \delta) K_{N-1}$ at $K_{N-1}$ and from (5.18), $m(f') \geq 0$, $M(f') < +\infty$. Remind $m(g') = 0$ and $M(g') = \lambda \delta$. Thus using iv) of the theorem 4, we have

$$0 < \partial_x u_{N-2}(x, T_{N-2}) < M(f') e^{-(\lambda+b)\Delta T_{N-2}} + \frac{\lambda \delta}{\lambda + b}(1 - e^{-(\lambda+b)\Delta T_{N-2}}) + \frac{(1-\delta) e^{-(\lambda+b)\Delta T_{N-2}}}{\sqrt{2\pi \overline{\sigma^2}(T_{N-2}, T_{N-2})}}, (\lambda + b > 0)$$

$$0 < \partial_x u_{N-2}(x, T_{N-2}) < M(f') + \frac{1-\delta}{\sqrt{2\pi \overline{\sigma^2}(T_{N-2}, T_{N-1})}}, \quad (\lambda = b = 0). \tag{5.27}$$

If $\delta = 1$, then $M(f') = 1$, $M(g') = \lambda$ from (5.25) and (5.19). Thus from (5.27) we have

$$0 < \partial_x u_{N-2}(x, T_{N-2}) < 1. \tag{5.28}$$

If $\delta < 1$ and the condition (5.14) holds, then from (5.20) we have $M(f') \leq d_{N-1}$ and thus from (5.27) we have

$$0 < \partial_x u_{N-2}(x, T_{N-2}) < d_{N-2} < 1. \tag{5.29}$$

(Again note that (5.14) is a sufficient condition for (5.28) to be true.) Using the theorem 5, we can obtain $\partial_x^2 u_{N-2}(x, T_{N-2})$ and finite number of convex or concave intervals of $u_{N-2}(x, T_{N-2})$ as the case when i=N−1 but here we will omit the detail.

Now let find the representation of $u_{N-2}(x,t)$ (the solution to the problem (5.21) and (5.25)). From (5.16), we can rewrite (5.25) as

$$u_{N-2}(x, T_{N-1}) = \bar{c}_N B_{K_N}^+(x, T_{N-1}; T_N) \cdot 1\{x > K_{N-1}\} + \delta A_{K_N}^-(x, T_{N-1}; T_N) \cdot 1\{x > K_{N-1}\} +$$
$$+ \lambda \int_{T_{N-1}}^{T_N} [\Phi_{N-2} B_{\Phi_{N-2}/\delta}^+(x, T_{N-1}; \tau) \cdot 1\{x > K_{N-1}\} + \delta A_{\Phi_{N-2}/\delta}^-(x, T_{N-1}; \tau) \cdot 1\{x > K_{N-1}\}] d\tau +$$
$$+ \bar{c}_{N-1} \cdot 1\{x > K_{N-1}\} + \delta x \cdot 1\{x \leq K_{N-1}\}.$$

Using the theorem 1, we have:

$$u_{N-2}(x,t) = e^{-\lambda(T_{N-1}-t)} \int_0^\infty \frac{e^{-\frac{\left(\ln\frac{x}{z} - b(T_{N-1}-t) - \frac{1}{2}\overline{\sigma^2}(t,T_{N-1})\right)^2}{2\overline{\sigma^2}(t,T_{N-1})}}}{z\sqrt{2\pi\overline{\sigma^2}(t,T_{N-1})}} u_{N-2}(z, T_{N-1}) dz +$$

$$+ \lambda \int_t^{T_{N-1}} e^{-\lambda(\tau-t)} \int_0^\infty \frac{e^{-\frac{\left(\ln\frac{x}{z} - b(\tau-t) - \frac{1}{2}\overline{\sigma^2}(t,\tau)\right)^2}{2\overline{\sigma^2}(t,\tau)}}}{z\sqrt{2\pi\overline{\sigma^2}(t,\tau)}} (\Phi_{N-2} 1\{z > \Phi_{N-2}/\delta\} + \delta \cdot z \cdot 1\{z < \Phi_{N-2}/\delta\}) dz d\tau.$$

From the definition of *second order binaries* [4, 16, 17], the lemma 2 of [17] on the integrals of binaries, we have

$$u_{N-2}(x, t) = \bar{c}_N B_{K_{N-1}K_N}^{++}(x, t; T_{N-1}, T_N) + \delta \cdot A_{K_{N-1}K_N}^{+-}(x, t; T_{N-1}, T_N) +$$
$$+ \lambda \int_{T_{N-1}}^{T_N} \left[\Phi_{N-1} B_{K_{N-1}\Phi_{N-1}/\delta}^{++}(x, t; T_{N-1}, \tau) + \delta \cdot A_{K_{N-1}\Phi_{N-1}/\delta}^{+-}(x, t; T_{N-1}, \tau)\right] d\tau + \quad (5.30)$$
$$+ \bar{c}_{N-1} B_{K_{N-1}}^+(x, t; T_{N-1}) + \delta \cdot A_{K_{N-1}}^-(x, t; T_{N-1})$$
$$+ \lambda \int_t^{T_{N-1}} \left[\Phi_{N-2} B_{\Phi_{N-2}/\delta}^+(x, t; \tau) + \delta \cdot A_{\Phi_{N-2}/\delta}^-(x, t; \tau)\right] d\tau, \quad T_{N-2} < t \leq T_{N-1}, \, x > 0.$$

Here $B_{a_1a_2}^{++}(x,t;T_1,T_2)$ and $A_{a_1a_2}^{+-}(x,t;T_1,T_2)$ are the prices of second order bond and asset binary options with the coefficients $r(t) \equiv \lambda$, $q(t) \equiv \lambda + b$, $\sigma(t)$, two expiry dates $T_1$ and $T_2$, two exercise prices $a_1$ and $a_2$, which are provided by the theorem 1 of [17, at page 5~6]. Returning to original variables using (5.15), we obtain the representation of $B_{N-2}$:

$$B_{N-2}(V, r, t) = Z(r, t; T) \cdot u_{N-2}(V/Z(r, t; T), t), \quad (T_{N-2} \leq t < T_{N-1}, \, x > 0)$$

Now let consider the case when $i=N-3$. In (5.5) and (5.7), we use (5.15) with $i=N-3$ and the notation (5.8) to get

$$\frac{\partial u_{N-3}}{\partial t} + \frac{1}{2}\sigma^2(t)x^2\frac{\partial^2 u_{N-3}}{\partial x^2} - bx\frac{\partial u_{N-3}}{\partial x} - \lambda u_{N-3} + \lambda \min\{\Phi_{N-3}, \delta x\} = 0, T_{N-3} < t < T_{N-2}, x > 0, \quad (5.31)$$

$$u_{N-3}(x, T_{N-2}) = [u_{N-2}(x, T_{N-2}) + \bar{c}_{N-2}] \cdot 1\{x > u_{N-2}(x, T_{N-2}) + \bar{c}_{N-2}\}$$
$$+ \delta x \cdot 1\{x \leq u_{N-2}(x, T_{N-2}) + \bar{c}_{N-2}\}, \, x > 0. \quad (5.32)$$

The equation for $u_{N-3}$ is just the same type of Black-Scholes equation for $u_{N-1}$ or $u_{N-2}$ but the terminal condition (5.32) is *not a standard* binary type. To make it into a standard binary type as the above, we consider the equation $x = u_{N-2}(x, T_{N-2}) + \bar{c}_{N-2}$. If (5.28) holds, then the equation has unique root $K_{N-2} \geq 0$ and we have $x > u_{N-2}(x,T_{N-2}) + \bar{c}_{N-2} \Leftrightarrow x > K_{N-2}$. Even though (5.28) does not hold, from the above considered properties of $u_{N-2}(x,T_{N-2})$ the set $S = \{x > 0: \partial_x u_{N-2}(x,T_{N-2}) = 1\}$ is a *finite* set and we can prove the following lemma:

**Lemma 2**. *The equation* $x = u_{N-2}(x,T_{N-2}) + \bar{c}_{N-2}$ *has unique root if and only if one of the two conditions satisfies.*

  (i) $u_{N-2}(x_1,T_{N-2}) + \bar{c}_{N-2} < x_1$ for every $x_1 \in S$.
  (ii) $u_{N-2}(x_1,T_{N-2}) + \bar{c}_{N-2} > x_1$ for every $x_1 \in S$.

Similarly as noted in the remark 8, in this lemma, (ii) seems not appropriate in financial meaning. But (i) seems to provide an appropriate (in financial meaning) upper bound for the coupon.

Thus, if $\bar{c}_{N-2}$ is in an appropriate range, then the equation has unique root $K_{N-2} \geq 0$ and we have $x > u_{N-2}(x,T_{N-2}) + \bar{c}_{N-2} \Leftrightarrow x > K_{N-2}$, too. Therefore the terminal value of the problem (5.32) can be written as follows:

$$u_{N-3}(x, T_{N-2}) = [u_{N-2}(x, T_{N-2}) + \bar{c}_{N-2}] \cdot 1\{x > K_{N-2}\} + \delta x \cdot 1\{x \leq K_{N-2}\}. \quad (5.33)$$

This gives the default barrier $K_{N-2}$ at time $T_{N-2}$.

The problem (5.31) and (5.33) is just the same problem for the inhomogenous Black-Scholes equation with binary type of terminal value. Thus using the theorem 1 in the section 2, the definition of *third order binaries* [16, 17] and the lemma 2 of [17] on the integrals of binaries, we have

$$\begin{aligned}
u_{N-3}(x,t) = &\, \bar{c}_N B^{+\,+\,+}_{K_{N-2}K_{N-1}K_N}(x,t;T_{N-2},T_{N-1},T_N) + \delta \cdot A^{+\,+\,-}_{K_{N-2}K_{N-1}K_N}(x,t;T_{N-2},T_{N-1},T_N) + \\
&+ \bar{c}_{N-1} B^{+\,+}_{K_{N-2}K_{N-1}}(x,t;T_{N-2},T_{N-1}) + \delta A^{+\,-}_{K_{N-2}K_{N-1}}(x,t;T_{N-2},T_{N-1}) \\
&+ \bar{c}_{N-2} B^{+}_{K_{N-2}}(x,t;T_{N-2}) + \delta \cdot A^{-}_{K_{N-2}}(x,t;T_{N-2}) \\
&+ \lambda \Bigg\{ \int_{T_{N-1}}^{T_N} [\Phi_{N-1} B^{+\,+\,+}_{K_{N-2}K_{N-1}\Phi_{N-1}/\delta}(x,t;T_{N-2},T_{N-1},\tau) + \delta \cdot A^{+\,+\,-}_{K_{N-2}K_{N-1}\Phi_{N-1}/\delta}(x,t;T_{N-2},T_{N-1},\tau)] d\tau \\
&+ \int_{T_{N-2}}^{T_{N-1}} [\Phi_{N-2} B^{+\,+}_{K_{N-2}\Phi_{N-2}/\delta}(x,t;T_{N-2},\tau) + \delta A^{+\,-}_{K_{N-2}\Phi_{N-2}/\delta}(x,t;T_{N-2},\tau)] d\tau \\
&+ \int_{t}^{T_{N-2}} [\Phi_{N-3} B^{+}_{\Phi_{N-3}/\delta}(x,t;\tau) + \delta \cdot A^{-}_{\Phi_{N-3}/\delta}(x,t;\tau)] d\tau \Bigg\}, \quad (T_{N-3} < t \leq T_{N-2},\ x > 0).
\end{aligned}$$

Returning to original variables using (5.15), we obtain the representation of $B_{N-3}$ :

$$B_{N-3}(V, r, t) = Z(r,t;T) \cdot u_{N-3}(V/Z(r,t;T),t),\quad (T_{N-3} < t \leq T_{N-2},\ x > 0).$$

Similarly (by induction), if we assume that we have already $B_{i+1}(V,r,t)$ ($T_{i+1} < t \leq T_{i+2}$), we can prove the following lemma:

**Lemma 3**. *Under the assumption of the theorem 6, if all the equations* $x = u_{i+1}(x,T_{i+1}) + \bar{c}_{i+1}$ ($i = 0,\cdots,N-2$) *have unique roots* $K_{i+1} \geq 0$, *then* $x > u_{i+1}(x,T_{i+1}) + \bar{c}_{i+1} \Leftrightarrow x > K_{i+1}$ *and furthermore the solution of (5.5) and (5.7) is provided as follows:*

$$B_i(V,r,t) = Z(r,t;T_N) \cdot u_i(V/Z(r,t;T_N),t), T_i < t \leq T_{i+1}.$$

*Here*

$$u_i(x,t) = \sum_{m=i}^{N-1}\left[\bar{c}_{m+1}B^{+\cdots++}_{K_{i+1}\cdots K_{m+1}}(x,t;T_{i+1},\cdots,T_{m+1}) + \delta\cdot A^{+\cdots+-}_{K_{i+1}\cdots K_m K_{m+1}}(x,t;T_{i+1},\cdots,T_m,T_{m+1})\right]$$

$$+ \lambda\sum_{m=i+1}^{N-1}\int_{T_m}^{T_{m+1}}\left[\Phi_m B^{+\cdots++}_{K_{i+1}\cdots K_m\frac{\Phi_m}{\delta}}(x,t;T_{i+1},\cdots,T_m,\tau) + \delta\cdot A^{+\cdots+-}_{K_{i+1}\cdots K_m\frac{\Phi_m}{\delta}}(x,t;T_{i+1},\cdots,T_m,\tau)\right]d\tau$$

$$+ \lambda\int_{t}^{T_{i+1}}\left[\Phi_i B^{+}_{\Phi_i/\delta}(x,t;\tau) + \delta\cdot A^{-}_{\Phi_i/\delta}(x,t;\tau)\right]d\tau,\quad T_i < t \le T_{i+1},\ x>0.$$

$B^{+\cdots+}_{K_1\cdots K_{m+1}}(x,t;T_1,\cdots,T_{m+1})$, $A^{+\cdots+-}_{K_1\cdots K_m K_{m+1}}(x,t;T_1,\cdots,T_m,T_{m+1})$ *are the price of m-th order bond and asset binaries with risk free rate $\lambda$, dividend rate $\lambda+b$ and volatility $\sigma(t)$* (the theorem 1 of [17, pages 5~6]).

Thus the theorem 6 has been proved. (QED)

Let denote the *k*-th *coupon rate* by $c_k = C_k/F$ ($k = 1,\ldots, N$). Then we have the following representation of the initial price of the our defaultable discrete coupon bond in terms of the debt, coupon rates, default recovery rate, default intensity, initial prices of the default free zero coupon bond and firm value and multi-variate normal distribution functions.

**Corollary 1.** *Under the assumption of theorem 1, the initial price of the bond can be represented as follows*:

$$B_0 = B_0(F, c_1, \cdots, c_N; \delta, \lambda; Z_0, V_0) =$$

$$= FZ_0\left[(1+c_N)e^{-\lambda T_N}N_N(d_1^-,\cdots,d_{N-1}^-,d_N^-;A_N) + \sum_{m=0}^{N-2}c_{m+1}e^{-\lambda T_{m+1}}N_{m+1}(d_1^-,\cdots,d_{m+1}^-;A_{m+1})\right]$$

$$+ \delta V_0\sum_{m=0}^{N-1}e^{-(\lambda+b)T_{m+1}}N_{m+1}(d_1^+,\cdots,d_m^+,-d_{m+1}^+;A_{m+1}) +$$

$$+ \lambda\sum_{m=0}^{N-1}\int_{T_m}^{T_{m+1}}[\,\delta V_0 e^{-(\lambda+b)\tau}N_{m+1}(d_1^+,\cdots,d_m^+,-\tilde{d}_{m+1}^+(\tau,\delta);\tilde{A}_{m+1}^-(\tau)) +$$

$$+ F(1+\Sigma_{k=1}^N c_k)Z_0 e^{-\lambda\tau}N_{m+1}(d_1^-,\cdots,d_m^-,\tilde{d}_{m+1}^-(\tau,\delta);\tilde{A}_{m+1}^-(\tau))]d\tau. \tag{5.34}$$

*Here* $N_m(a_1,\cdots,a_m;A)$ *is the* cumulative distribution function *of m-variate normal distribution* with *zero mean* vector and a *covariance* matrix $A^{-1}$ (the theorem 1 of [17]),

$$N_m(a_1,\cdots,a_m;A) = \int_{-\infty}^{a_1}\cdots\int_{-\infty}^{a_m}\frac{1}{(\sqrt{2\pi})^m}\sqrt{\det A}\exp(-\frac{1}{2}y^\perp A y)dy,$$

$$d_i^\pm = \left(\int_0^{T_i}\sigma^2(t)dt\right)^{-1/2}\left(\ln\frac{V_0}{Z_0 K_i} - bT_i \pm \frac{1}{2}\int_0^{T_i}\sigma^2(t)dt\right), i = 1,\cdots, N-1;$$

$$d_N^\pm = \left(\int_0^{T_N}\sigma^2(t)dt\right)^{-1/2}\left(\ln\frac{V_0}{FZ_0(1+c_N)} - bT_N \pm \frac{1}{2}\int_0^{T_N}\sigma^2(t)dt\right);$$

$$\tilde{d}_i^{\pm}(\tau,\delta) = \left(\int_0^{\tau}\sigma^2(t)dt\right)^{-1/2}\left(\ln\frac{\delta V_0}{Z_0 F(1+\Sigma_{k=1}^N c_k)} - b\tau \pm \frac{1}{2}\int_0^{\tau}\sigma^2(t)dt\right), T_{i-1} \leq \tau < T_i; i=1,\cdots,N.$$

*The matrix* $(A_m)^{-1} = (r_{ij})_{i,j=1}^m$ *is given by*

$$r_{ij} = \sqrt{\int_0^{T_i}\sigma^2(t)dt \Big/ \int_0^{T_j}\sigma^2(t)dt}, \quad r_{ji} = r_{ij}, \ i \leq j \ (i,j=1,\cdots,m),$$

$(\tilde{A}_m(\tau))^{-1} = (\tilde{r}_{ij}(\tau))_{i,j=1}^m$ *is the matrix whose m-th row and column are given by*

$$\tilde{r}_{im}(\tau) = \sqrt{\int_0^{T_i}\sigma^2(t)dt \Big/ \int_0^{\tau}\sigma^2(t)dt}, \quad \tilde{r}_{mi}(\tau) = \tilde{r}_{im}(\tau), \ i < m \ (i=1,\cdots,m-1)$$

*and other elements coincide with those of* $(A_m)^{-1}$. *The matrices* $(A_m^-)^{-1} = (r_{ij}^-)_{i,j=1}^m$ *and* $(\tilde{A}_m^-(\tau))^{-1}$

$= (\tilde{r}_{ij}^-(\tau))_{i,j=1}^m$ *have such m-th rows and columns that*

$$r_{im}^- = -r_{im}, \quad r_{mi}^- = -r_{mi}; \quad \tilde{r}_{im}^-(\tau) = -\tilde{r}_{im}(\tau), \quad \tilde{r}_{mi}^-(\tau) = -\tilde{r}_{mi}(\tau), \ i < m \ (i=1,\cdots,m-1)$$

*and other elements coincide with those of* $(A_m)^{-1}$ *and* $(\tilde{A}_m(\tau))^{-1}$, *respectively*.

> **Remark 10.** If $b = 0$ and $\lambda = 0$, then our pricing formula (5.34) nearly coincides with the formula (5) of [1, at page 752] and the only difference comes from the fact that $k$-th coupon is provided as a discounted value of the maturity-value in our model. If $b = 0$, $C_k = 0 (i = 1,\ldots, N)$ and $\lambda = 0$, then the formula (5.34) coincides with the formula (5) of [1, at page 752] when $C_k = 0$ ($i = 1,\ldots, N$) which is a known formula for defaultable zero coupon bond that generalizes Merton (1974) [15].